\documentclass[pre,showpacs,preprint]{revtex4}
\usepackage{bm}
\usepackage{amsmath}

\begin{document}

\title{\textbf{Kinetic Theory of Response Functions for the Hard Sphere
Granular Fluid}}
\author{Aparna Baskaran}
\affiliation{Physics Department, Syracuse University, Syracuse, NY 13244}
\author{James W. Dufty}
\affiliation{Department of Physics, University of Florida, Gainesville, FL 32611}
\author{ J. Javier Brey}
\affiliation{F\'{\i}sica Te\'{o}rica, Universidad de Sevilla, Apartado de Correos 1065,
E-41080, Sevilla, Spain}
\date{\today }

\begin{abstract}
The response functions for small spatial perturbations of a homogeneous
granular fluid have been described recently. In appropriate dimensionless
variables, they have the form of stationary state time correlation
functions. Here, these functions are expressed in terms of reduced single
particle functions that are expected to obey a linear kinetic equation. The
functional assumption required for such a kinetic equation, and a Markov
approximation for its implementation are discussed. If, in addition, static
velocity correlations are neglected, a granular fluid version of the
linearized Enskog kinetic theory is obtained. The derivation makes no \emph{%
a priori} limitation on the density, space and time scale, nor degree of
inelasticity. As an illustration, recently derived Helfand and Green-Kubo
expressions for the Navier-Stokes order transport coefficients are evaluated
with this kinetic theory. The results are in agreement with those obtained
from the Chapman-Enskog solution to the nonlinear Enskog kinetic equation.
\end{abstract}

\maketitle

%\date{\today }

\section{Introduction}

One of the most productive methods to study transport in normal fluids is
through the measurement, simulation, and theory of linear response functions
\cite{Martin68,Forster75,Boon91,McL89}. Of particular interest are those
that describe the linear response of the ``hydrodynamic fields'' (mass,
energy, and momentum densities) to small spatial perturbations of the
homogeneous equilibrium state. The terminology, hydrodynamic fields, is due
to the fact that these are the variables expected to obey the
phenomenological hydrodynamic equations on large space and time scales. The
response functions provide the means to study such hydrodynamic excitations
starting from their fundamental basis in non-equilibrium statistical
mechanics. For example, they provide the formally exact Helfand \cite%
{Helfand60} and Green-Kubo \cite{McL63,McL89} representations for transport
coefficients. More generally, the response functions describe the broader
range of excitations on shorter space and time scales as well. One of the
most instructive theoretical approaches to their evaluation has been kinetic
theory, with the greatest progress made for the idealized fluid of hard
spheres \cite{Resibois77,McL89}.

Recently, this linear response approach has been extended to granular fluids
\cite{DBB06,BDB06}. The objective here is to demonstrate the application of
kinetic theory methods for the evaluation of the granular response
functions. Only the case of smooth, inelastic hard spheres is considered
both for simplicity and to parallel closely the corresponding developments
for normal fluids. Such an idealized model still captures the most important
features of many granular fluids \cite{chemE}. The usual notion of kinetic
theory is a nonlinear equation for the probability density in single
particle phase space. An advantage of the linear response functions is that
their kinetic equation is inherently linear. For practical purposes, a
Markovian approximation to this linear kinetic equation is described, based
on the neglect of dynamical correlations. The approximations leading to this
equation, and the differences between its implications for normal and
granular fluids, are discussed. The nature of the approximation does not
\textit{a priori} assume weak dissipation, low density, or large length and
time scales. In the elastic limit, it becomes the linear Enskog kinetic
equation for the response functions of a normal equilibrium fluid. A
granular Enskog limit is described here as well, by the further neglect of
all velocity correlations in the Markov approximation.

A related set of time correlation functions involving the fluxes of the
hydrodynamic fields, instead of the fields themselves, determine the Helfand
and Green-Kubo representations for the transport coefficients of the
Navier-Stokes order hydrodynamic equations \cite{DBB06,BDB06}. These can be
evaluated by the same kinetic theory developed for the response functions.
This is described in detail for the shear viscosity in the Markov
approximation, indicating how the time dependence of these flux correlation
functions can be determined, as well as the associated transport
coefficient. The remaining Navier-Stokes transport coefficients are
evaluated in the Appendix \ref{ap4}. It is confirmed that, in the linear
Enskog approximation, the results agree in detail with those obtained from
the Chapman-Enskog method to solve the nonlinear Enskog equation for the
distribution function \cite{GD99}.

A similar program has already been carried out in the limited context of
dynamics for an impurity particle in a granular fluid \cite{DBL02,GD01}. In
that case, the only hydrodynamic field is the impurity particle probability
density, and the response function is its autocorrelation function. The
hydrodynamic equation is a diffusion equation, and the Green-Kubo expression
for the diffusion coefficient is given by the time integral of the velocity
autocorrelation function. This was evaluated by kinetic theory in the Enskog
approximation, and both the correlation function and the transport
coefficient were compared with molecular dynamics simulation data over a
wide range of densities and degrees of inelasticity. The results provide an
instructive characterization of the domain of validity for the Enskog
(Enskog-Lorentz in this case) kinetic equation, and expose important
differences between impurity dynamics in normal and granular fluids. The
presentation here constitutes an extension of that theoretical analysis to
the full range of multi-particle mass, energy, and momentum transport.

The origin of a kinetic theory is an exact hierarchy (the BBGKY hierarchy
\cite{Resibois77,McL89}) of equations for the reduced few particle
representations for a property of interest. A kinetic equation is comprised
of the first hierarchy equation together with a ``closure'', expressing the
solution to the second hierarchy equation as a functional of that for the
first. This leads to a closed, deterministic equation for the latter which
is the kinetic equation. Practical methods have been developed for normal
fluids based on inversion of cluster expansions and partial resummations, as
well as more phenomenological estimates. This is the point at which kinetic
theory confronts the difficult many body problem, and one objective of the
current work is to motivate a corresponding attention to such details for
the granular fluid. Only in this way can the qualitative speculations about
differences between normal and granular fluid be made more precise. An
example of such uncertainties is the role of velocity correlations in the
construction of the closure. It is well known that velocity correlations
generated by collective many particle collisions (e.g., ring collisions) are
responsible for the dominant density dependence of transport coefficients at
very high densities, but they are relatively unimportant at low to moderate
densities where simple spatial correlations (e.g., excluded volume effects)
are dominant. The latter are incorporated in the Enskog approximation for
accurate corrections to the predictions of the low density Boltzmann
equation. Granular fluids introduce a complication to this separation of
dynamical velocity correlations and static structural correlations,
according to the density considered. In these systems, there are inherent
static velocity correlations, not directly associated with many particle
dynamics, that are present even at low to moderate densities. The kinetic
theory for response functions provides an appropriate setting for the study
of the quantitative importance of these correlations on properties of
interest.

The response functions in the Markovian approximation are expressed in terms
of the linear generator for dynamics. Questions about the existence and
dominance of hydrodynamics can be made precise at this point, by asking if
the hydrodynamic modes (eigenvalues) appear in the spectrum of this operator
and if they are the slowest modes. The first part, existence of hydrodynamic
modes, can be demonstrated at long wavelengths if expected conditions of
analyticity are satisfied (see Sec. \ref{s6} below). The second issue of
dominance at long times can be addressed practically using simplified
kinetic models for this generator. Such results support the primary
assumptions in references \cite{DBB06} and \cite{BDB06}, for a derivation of
formally exact expressions for the hydrodynamic transport coefficients in
terms of time correlation functions. However, the linear kinetic equation
applies as well to short space and time scales, that are important for
non-hydrodynamic response.

The main points of this analysis are summarized in the last section.

\section{Linear Response Functions}

\label{s2}

An idealized granular fluid of $N$ smooth, inelastic hard spheres ($d=3$) or
disks ($d=2$) of mass $m$ and diameter $\sigma$ is considered. The
inelasticity is characterized by a constant coefficient of normal
restitution $\alpha $. The properties of interest are the average number
density $n({\bm r},t)$, the granular temperature $T({\bm r},t)$, and the
local flow velocity ${\bm
U}({\bm r},t)$, collectively denoted by $y({\bm r},t) \equiv \left\{
y_{\beta}({\bm r},t) \right\}$. The response to be studied here is the
variation of these fields at time $t$ due to a variation in their initial
values at time $t=0$. The initial conditions are spatial variations of a
homogeneous reference state, $y_{\beta }({\bm r},0)=y_{\beta , h}(0)+\delta
y_{\beta }({\bm r},0)$. In the most general case, the response $\delta
y_{\beta }[{\bm r},t | y \left( 0\right)]$ depends nonlinearly on the
perturbations $\delta y({\bm r},0)$, but if these are small, the linear
order gives the dominant response,
\begin{equation}
\delta y_{\beta }\left[ {\bm r},t | y \left( 0\right) \right] =
\sum_{\gamma} \int d{\bm r}^{\prime }\, C_{\beta \gamma }\left( {\bm r}-{\bm %
r}^{\prime },t\right) \delta y_{\gamma }({\bm r}^{\prime },0),  \label{2.1}
\end{equation}
with the linear response functions defined as
\begin{equation}
C_{\beta \gamma }\left( {\bm r}-{\bm r}^{\prime },t\right) =\left[\frac{%
\delta y_{\beta }\left[{\bm r},t|y(0)\right]}{\delta y_{\gamma }({\bm r}%
^{\prime },0)}\right] _{\delta y(0)= 0 }.  \label{2.2}
\end{equation}
Since the reference state is homogeneous, the linear response functions
depend on ${\bm r}$ and ${\bm r}^{\prime }$ only through their difference.

The difference between granular and normal fluids occurs already at the
level of the homogeneous reference state. For normal fluids, this is the
equilibrium stationary state and all the time dependence of the response
functions is due to the spatial perturbations. For granular fluids, the
homogeneous reference state is inherently time-dependent, even without
perturbation, due to the ``cooling'' of inelastic collisions. As a
consequence, the temperature $T_{h}$ of the homogeneous granular fluid
decreases in time according with a cooling law
\begin{equation}
\frac{\partial T_{h}(t)}{\partial t}=-\zeta _{0}\left[ T_{h}(t)\right]
T_{h}(t),  \label{2.3}
\end{equation}
where the cooling rate $\zeta _{0}\left( T_{h}\right) $ is a characteristic
function of the homogeneous cooling reference state. Thus the relevant
response at time $t$ is measured relative to the reference homogeneous state
at the same time rather than to the initial state. Then, dimensionless
fields $\delta y^{*}_ {\beta}$ are introduced by
\begin{equation}
\left\{ \delta y_{\beta}^{*} \right\} \equiv \left\{ \frac{\delta y_{\beta}}{%
\overline{y}_{\beta,h}(t)} \right\} \equiv \left\{ \frac{\delta n}{n_{h}},
\frac{\delta T }{T_{h}(t)},\frac{\delta {\bm U}}{v_{0}(t)} \right\},
\label{2.4}
\end{equation}
where the definition of the reference fields $\overline{y}_{\beta,h} (t)$
follows from the second identity and
\begin{equation}
v_{0}(t) \equiv \left[ \frac{2T_{h}(t)}{m} \right]^{1/2}  \label{2.5}
\end{equation}
is a thermal velocity. The dependence of the linear response functions on ${%
\bm r}-{\bm r}^{\prime}$ suggests the utility of a Fourier representation of
Eq.\ (\ref{2.1}) that is expressed in the form
\begin{equation}
\delta \widetilde{y}_{\beta }^{*}({\bm k}^{*},s)=\sum_{\gamma} \widetilde{C}%
_{\beta \gamma }^{*} \left( {\bm k}^{*},s\right) \delta \widetilde{y}%
_{\gamma }^{*}({\bm k}^{*},0),  \label{2.6}
\end{equation}
where a tilde over a function denotes its dimensionless Fourier transform,
defined by
\begin{eqnarray}
\widetilde{f}({\bm k}^{*}) & \equiv & \ell^{-d} \int d{\bm r}\, e^{i{\bm %
k^{*}\cdot {\bm r}}/\ell }f\left( {\bm r}\right)  \notag \\
&=& \int d{\bm r}^{*} e^{i {\bm k}^{*} \cdot {\bm r}^{*}} f\left( {\bm r}%
\right),  \label{2.7}
\end{eqnarray}
with ${\bm r}^{*} ={\bm r}/\ell$, and dimensionless response functions have
been identified as
\begin{eqnarray}
\widetilde{C}_{\beta \gamma }^{\ast }\left( {\bm k}^{\ast },s\right) & = &
\overline{y}_{\beta, h}^{-1}(t) \ell^{d} \widetilde{C}_{\beta \gamma} ({\bm k%
}^{*},t) \overline{y}_{\gamma,h}(0)  \notag \\
& = &\overline{y}_{\beta, h}^{-1}(t) \int d{\bm r}\, e^{i{\bm k}^{*} \cdot {%
\bm r} /\ell }C_{\beta \gamma }\left({\bm r},t\right) \overline{y}_{\gamma,
h}(0).  \label{2.8}
\end{eqnarray}
Here ${\bm k}^{\ast }={\bm k}\ell $ is a dimensionless wavevector, with $%
\ell $ being a characteristic length of the system. Moreover, a
dimensionless time scale
\begin{equation}
s= \int_{0}^{t} dt^{\prime}\, \frac{v_{0}(t^{\prime})}{\ell }  \label{2.9}
\end{equation}
has been introduced. This time $s$ has the interpretation of an average
collision number per particle up to time $t$, if $\ell$ is chosen to be the
mean free path of the particles. The remainder of this presentation focuses
on the dimensionless linear response functions $\widetilde{C}_{\beta
\gamma}^{*}$.

For small ${\bm k}^{\ast }$ and large $s$, the functions $\delta y_{\beta
}^{\ast }( {\bm k}^{\ast },s)$ are expected to obey the linearized
hydrodynamic Navier-Stokes equations, and the response functions correspond
to the Green functions for the solution to the initial value problem
associated with those equations \cite{BDB06}. That description is only
phenomenological, since it is parameterized by the unknown transport
coefficients. A more complete and exact description for all ${\bm k}^{\ast }$
and $s$ is provided by non-equilibrium statistical mechanics, as described
in references \cite{DBB06} and \cite{BDB06}. Briefly, the construction is as
follows.

The hydrodynamic fields $y_{\beta}({\bm r},t)$ are defined from averages of
the microscopic number density, energy density, and momentum density over
the phase space density $\rho (\Gamma,t)$, representing the probability that
the positions ${\bm q}_{r}$ and velocities ${\bm v}_{r}$ of the particles
have specified values denoted by $\Gamma \equiv \left\{ x_{1}, \ldots, x_{N}
\right\}$. The latter is a point in the $2Nd$ dimensional phase space with
the notation $x_{r} \equiv \left\{ {\bm q}_{r}, {\bm v}_{r} \right\}$. For
any specified initial state, $\rho (\Gamma,0)$, this probability density
evolves in time according to the Liouville equation
\begin{equation}
\left( \frac{\partial}{\partial t} + \overline{L} \right) \rho (\Gamma,t) =
0,  \label{2.10}
\end{equation}
with
\begin{equation}
\overline{L} = \sum_{r=1}^{N} {\bm v}_{r} \cdot \frac{\partial}{\partial {%
\bm q}_{r}}-\frac{1}{2}\sum_{r=1}^{N} \sum_{s \neq r}^{N} \overline{T}
(x_{r},x_{s}).  \label{2.11}
\end{equation}
The time independent operator $\overline{L}$ is the generator for the hard
sphere dynamics, where the singular binary collisions are described by
\begin{equation}
\overline{T}(x_{r},x_{s})=\delta (q_{rs}-\sigma )|\widehat{\bm q}_{rs} \cdot
{\bm g}_{rs} |\left[ \Theta \left( \widehat{\bm q}_{rs} \cdot {\bm g}_{rs}
\right) \alpha^{-2} b_{rs}^{-1}-\Theta \left( - \widehat{\bm q}_{rs} \cdot {%
\bm g}_{rs} \right) \right] .  \label{2.12}
\end{equation}
In this expression, ${\bm q}_{rs} = {\bm q}_{r}-{\bm q}_{s}$, ${\bm
g}_{rs} = {\bm v}_{r}-{\bm v}_{s}$, $\Theta$ is the Heaviside step function,
$\widehat{\bm q}_{rs} \equiv {\bm q}_{rs}/ q_{rs}$, and $b_{rs}^{-1}$ is the
substitution operator that replaces the velocities ${\bm v}_{r}$, ${\bm v}%
_{s}$ by their ``precollisional'' values ${\bm v}_{r}^{\prime \prime}$,${\bm %
v}_{s}^{ \prime \prime }$,
\begin{equation}
b_{rs}^{-1}F\left({\bm v}_{r},{\bm v}_{s}\right) =F\left( {\bm v}_{r}^{
\prime \prime },{\bm v}_{s}^{ \prime \prime }\right) ,  \label{2.13}
\end{equation}
\begin{equation}
{\bm v}_{r}^{ \prime \prime }={\bm v}_{r} -\frac{ 1+\alpha }{2\alpha }\left(
\widehat{\bm q}_{rs} \cdot {\bm g}_{rs}\right) \widehat{\bm q}_{rs},
\label{2.14}
\end{equation}
\begin{equation}
{\bm v}_{s}^{\ast \prime \prime }={\bm v} _{s}+ \frac{ 1+\alpha }{2\alpha }%
\left( \widehat{\bm q}_{rs} \cdot {\bm g}_{rs} \right) \widehat{\bm q}_{rs}.
\label{2.15}
\end{equation}

For an isolated system, instead of the equilibrium state for a normal fluid,
there is a spacial solution to the Liouville equation of the form
\begin{equation}
\rho_{h}(\Gamma,t) = \left[ \ell v_{0}(t) \right]^{-Nd} \rho^{*}_{h} \left(
\left\{ {\bm q}_{rs}/\ell, {\bm v}_{r}/v_{0}(t) \right\} \right) \equiv %
\left[ \ell v_{0}(t) \right]^{-Nd} \rho^{*}_{h} (\Gamma^{*}).  \label{2.16}
\end{equation}
The dimensionless phase point $\Gamma^{*} \equiv \left\{ x^{*}_{1},
\ldots,x^{*}_{N} \right\}$ is now expressed in terms of the scaled positions
and velocities $x^{*}_{r} = \left\{ {\bm q}^{*}_{r},{\bm
v}^{*}_{r} \right\} \equiv \left\{ {\bm q}_{r} /\ell, {\bm
v}_{r}/v_{0}(t) \right\}$. This solution depends on the positions only
through the relative variables ${\bm q}_{rs}$ and, therefore, it has
translational invariance, representing a homogeneous state. All of the time
dependence occurs through the thermal velocity defined by Eqs. (\ref{2.5})
and (\ref{2.3}), so the Liouville equation becomes for $\rho_{h}^{*}(%
\Gamma^{*})$
\begin{equation}
\overline{\mathcal{L}}^{\ast }\rho _{h}^{\ast }(\Gamma^{*} )=0.  \label{2.17}
\end{equation}
The operator $\overline{\mathcal{L}}^{*}$ is the sum of the original
generator for trajectories, now in the dimensionless variables, plus a
scaling operator representing the time dependenc of $v_{0}(t)$ using Eq. (%
\ref{2.3}),
\begin{equation}
\overline{\mathcal{L}}^{\ast }=\overline{L}^{\ast } +\frac{\zeta _{0}^{\ast }%
}{2}\sum_{r=1}^{N} \frac{\partial}{\partial {\bm v}^{*}_{r}} \cdot {\bm v}%
_{r}^{*},  \label{2.18}
\end{equation}
where $\zeta_{0}^{*}$ is the dimensionless cooling rate
\begin{equation}
\zeta _{0}^{\ast } = \frac{\ell }{v_{0}(t)}\, \zeta _{0}\left[ T_{h}(t)%
\right] .  \label{2.19}
\end{equation}
The solution to Eq.\ (\ref{2.17}) will be referred to as the homogeneous
cooling state (HCS).

For more general solutions to the Liouville equation, it is useful to
introduce the same scaled velocities to account for this inherent cooling.
Then, the Liouville equation in dimensionless form becomes
\begin{equation}
\left( \frac{\partial}{\partial s} +\overline{\mathcal{L}}^{*} \right)
\rho^{*} (\Gamma^{*},s) =0.  \label{2.20}
\end{equation}
In this form, it is seen that the HCS is a stationary solution to the
dimensionless Liouville equation. This is an important result for the
representation of granular response functions. It shows that, in the
appropriate dimensionless form, the reference state is again stationary,
just as for equilibrium fluids. However, the introduction of this stationary
representation comes at the price of changing the generator for the dynamics
from $\overline{L}^{*}$ to $\overline{\mathcal{L}}^{*}$. For the purposes of
the discussion here, it is assumed that all properties of this homogeneous
reference state are known. Further comments on the HCS are given in the
Appendix \ref{ap1}.

The deviations of the relative hydrodynamic fields form their values in the
HCS, $\delta \widetilde{y}_{\beta }^{\ast }({\bm k}^{\ast },s)$, are the
averages of associated phase functions $\widetilde{a}_{\beta }^{\ast
}(\Gamma ^{\ast },s)$,
\begin{equation}
\delta \widetilde{y}_{\beta }^{\ast }({\bm k}^{\ast },s)=\int d\Gamma ^{\ast
}\,\widetilde{a}_{\beta }^{\ast }\left( \Gamma ^{\ast };{\bm k}^{\ast
}\right) \left[ \rho ^{\ast }(\Gamma ^{\ast },s)-\rho _{h}^{\ast }(\Gamma
^{\ast })\right] ,  \label{2.21}
\end{equation}%
with
\begin{equation}
\widetilde{a}_{\beta }^{\ast }\left( \Gamma ^{\ast };{\bm k}^{\ast }\right) =%
\frac{1}{n_{h}^{\ast }}\sum_{r=1}^{N}e^{i{\bm k}^{\ast }\cdot {\bm q}%
_{r}^{\ast }}a_{\beta }^{\ast }\left( {\bm v}_{r}^{\ast }\right) ,
\label{2.22}
\end{equation}%
where $n_{h}^{\ast }\equiv n_{h}\ell ^{d}$ is the number density of the HCS
in the reduced units and the single particle functions $a_{\beta }^{\ast }({%
\bm v^{\ast }}_{r})$ are defined by
\begin{equation}
\left\{ a_{\beta }^{\ast }({\bm v})\right\} \equiv \left\{ 1,\frac{2v^{\ast
2}}{d}-1,v_{\parallel }^{\ast },{\bm v}_{\perp }^{\ast }\right\} .
\label{2.23}
\end{equation}%
For later convenience, the components of the flow field and the velocity of
the particles have been chosen to be a longitudinal component along ${\bm k}%
^{\ast }$, and $d-1$ transverse components, so that $v_{\parallel }^{\ast
}\equiv \widehat{\bm k}\cdot {\bm v}^{\ast }$ and $v_{\perp i}^{\ast }\equiv
\widehat{\bm e}_{i}\cdot {\bm v}^{\ast }$, with $\left\{ \widehat{\bm k}%
\equiv {\bm k}^{\ast }/k^{\ast },\widehat{\bm e}_{i};i=1,\ldots ,d-1\right\}
$ forming a set of $d$ pairwise perpendicular unit vectors. To formulate the
linear response problem, the initial state $\rho ^{\ast }(\Gamma ^{\ast },0)$
for the solution to the Lioville equation in (\ref{2.20}) is chosen to be
close to the HCS, in the sense that it is a functional of the initial fields
$\delta y_{\beta }^{\ast }({\bm r}^{\ast },0)$ and becomes the HCS for $%
\delta y_{\beta }^{\ast }\rightarrow 0$. More specifically, the system is
viewed as partitioned into small cells such that the distribution function
is the HCS in each cell, but with different values for the hydrodynamic
fields. This is the analogue of the local equilibrium distribution for
normal fluids, and will be referred to as the \emph{local} HCS, $\rho
_{lh}^{\ast }[\Gamma ^{\ast }|\delta y^{\ast }]$. Its construction is
discussed in Appendix \ref{ap1}. For small $\delta y_{\beta }^{\ast }$, the
solution to the Liouville equation for the initial local HCS expanded to
first order is
\begin{eqnarray}
\rho ^{\ast }(\Gamma ^{\ast },s)-\rho _{h}^{\ast }(\Gamma ^{\ast }) &=&e^{-s%
\mathcal{L}^{\ast }}\rho ^{\ast }(\Gamma ^{\ast },0)  \notag \\
&\rightarrow &e^{-s\mathcal{L}^{\ast }}\sum_{\gamma =1}^{d+2}\int d{\bm r}%
^{\ast }\,\left[ \frac{\delta \rho _{\ell h}^{\ast }\left[ \Gamma ^{\ast
}|\delta y^{\ast }\right] }{\delta y_{\gamma }^{\ast }\left( {\bm r}^{\ast
}\right) }\right] _{\delta y^{\ast }=0}\delta y_{\gamma }^{\ast }({\bm r}%
^{\ast },0).  \label{2.24}
\end{eqnarray}%
Substitution into Eq.\ (\ref{2.21}) allows identification of the response
functions $C_{\beta \gamma }^{\ast }({\bm r}^{\ast },s)$ and their
equivalent Fourier representation (using translational invariance of the
functional derivative at $\delta y^{\ast }=0$),
\begin{equation}
\widetilde{C}_{\beta \gamma }^{\ast }\left( {\bm k},s\right) =\int d\Gamma
^{\ast }\,\widetilde{a}_{\beta }^{\ast }\left( \Gamma ^{\ast };{\bm k}^{\ast
}\right) \psi _{\gamma }^{\ast }\left( \Gamma ^{\ast };{\bm0}^{\ast
},s\right) ,  \label{2.25}
\end{equation}%
with
\begin{equation}
\psi _{\gamma }^{\ast }\left( \Gamma ^{\ast };{\bm r}^{\ast },s\right) =e^{-s%
\overline{\mathcal{L}}^{\ast }}\psi _{\gamma }^{\ast }\left( \Gamma ^{\ast };%
{\bm r}^{\ast }\right)  \label{2.26}
\end{equation}%
and
\begin{equation}
\psi _{\gamma }^{\ast }\left( \Gamma ^{\ast };{\bm r}^{\ast }\right) =\left[
\frac{\delta \rho _{\ell h}^{\ast }\left[ \Gamma ^{\ast }|\delta y^{\ast }%
\right] }{\delta y_{\gamma }^{\ast }\left( {\bm r}^{\ast }\right) }\right]
_{\delta y=0}.  \label{2.27}
\end{equation}%
These results are closely analogous to those for a normal fluid, where $\rho
_{\ell h}^{\ast }[\Gamma ^{\ast }|\delta y^{\ast }]$ is an initial local
equilibrium ensemble and $\rho _{h}^{\ast }(\Gamma ^{\ast })$ is the
reference equilibrium Gibbs ensemble, $\rho _{e}^{\ast }(\Gamma ^{\ast })$.
For example, if the grand canonical ensemble were used, the $\widetilde{\psi
}_{\gamma }^{\ast }$'s would become linear combinations of the set $\left\{
\widetilde{a}_{\gamma }^{\ast }\right\} $ times the equilibrium ensemble,
and the response functions would be equilibrium time correlation functions
for the local conserved quantities.

\section{Kinetic Theory}

\label{s3}

All of the following analysis is carried out in terms of the dimensionless
variables, so the asterisk will be left implicit for simplicity. The
response functions of Eq.\ (\ref{2.25}) are expressed in terms of the full $%
N $-particle phase space. A reduced description in terms of the single
particle phase space is possible because the functions $\widetilde{a}_{\beta
}\left( \Gamma ,{\bm k}\right) $ defined in Eq.\ (\ref{2.22}) are sums of
single particle functions. Consequently, integrating over the positions and
momenta for all except one particle leads to the exact alternative
representation:
\begin{equation}
\widetilde{C}_{\beta \gamma }\left( {\bm k},s\right) =\frac{1}{n_{h}}\int d{%
\bm v}_{1}\,a_{\beta }\left( {\bm v}_{1}\right) \widetilde{\psi }_{\gamma
}^{(1)}({\bm v}_{1},{\bm k},s),  \label{3.1}
\end{equation}%
where
\begin{equation}
\widetilde{\psi }_{\gamma }^{(1)}({\bm v}_{1},{\bm k},s)=\int d{\bm q}%
_{1}e^{i{\bm k}\cdot {\bm q}_{1}}\psi _{\gamma }^{(1)}(x_{1},s).  \label{3.3}
\end{equation}%
The function $\psi _{\gamma }^{(1)}(x_{1},s)$ is the first element of a
hierarchy of $N$ functions $\psi _{\gamma }^{(m)}(x_{1},x_{2},\ldots
,x_{m},s)$, $m=1,\ldots ,N$, defined through
\begin{equation}
\psi _{\gamma }^{(m)}(x_{1},\ldots ,x_{m},s)\equiv \left[ \frac{\delta
f^{(m)}(x_{1},\ldots ,x_{m},s)}{\delta y_{\gamma }({\bm0},0)}\right]
_{\delta y=0},  \label{3.3a}
\end{equation}%
\begin{equation}
f^{(m)}(x_{1},\ldots ,x_{m},s)\equiv \frac{N!}{(N-m)!}\int
dx_{m+1}..dx_{N}\,\rho (\Gamma ,s),  \label{3.3b}
\end{equation}%
$1\leq m\leq N$. The functions $f^{(m)}(x_{1},\ldots ,x_{m},s)$ are the
reduced distribution functions associated with the solution to the Liouville
equation (\ref{2.20}). They obey the corresponding BBGKY hierarchy of
equations \cite{McL89}. The first equation of this hierarchy is
\begin{equation}
\left( \frac{\partial }{\partial s}+{\bm v}_{1}\cdot \frac{\partial }{%
\partial {\bm q}_{1}}\right) f^{(1)}(x_{1},s)+\frac{\zeta _{0}}{2}\frac{%
\partial }{\partial {\bm v}_{1}}\cdot \left[ {\bm v}_{1}f^{(1)}(x_{1},s)%
\right] =\int dx_{2}\,\overline{T}(x_{1},x_{2})f^{(2)}(x_{1},x_{2},s).
\label{3.3c}
\end{equation}%
Then, it follows directly from the definition (\ref{3.3a}) that the function
$\psi _{\gamma }^{(1)}(x_{1},s)$ obeys the analogous equation
\begin{equation}
\left( \frac{\partial }{\partial s}+{\bm v}_{1}\cdot \frac{\partial }{%
\partial {\bm q}_{1}}\right) \psi _{\gamma }^{(1)}(x_{1},s)+\frac{\zeta _{0}%
}{2}\frac{\partial }{\partial {\bm v}_{1}}\cdot \left[ {\bm v}_{1}\psi
_{\gamma }^{(1)}(x_{1},s)\right] =\int dx_{2}\,\overline{T}(x_{1},x_{2})\psi
_{\gamma }^{(2)}(x_{1},x_{2},s).  \label{3.5}
\end{equation}

The representation given by Eq.\ (\ref{3.1}) is very appealing, since the $N$%
-particle problem has been expressed without approximation in terms of the
effective dynamics in the single particle phase space. The fundamental
difficulty, however, is that the first hierarchy equation (\ref{3.5}) does
not determine this effective dynamics without specifying $\psi _{\gamma
}^{(2)}(x_{1},x_{2},s)$. An equation similar to (\ref{3.5}) can be written
for $\psi _{\gamma }^{(2)}(x_{1},x_{2},s)$, the second BBGKY hierarchy
equation, but it in turn requires specification of $\psi _{\gamma
}^{(3)}(x_{1},x_{2},x_{3},s)$. In this way, a coupling to the full $N$
particle problem recurs. This coupling is broken if $\psi _{\gamma
}^{(2)}(x_{1},x_{2},s)$ can be specified as an explicit functional of $\psi
_{\gamma }^{(1)}(x_{1},s).$ It is argued in Appendix \ref{ap2} that this
functional, when it exists, is independent of $\delta y_{\gamma }({\bm r},0)$%
, linear in $\psi _{\gamma }^{(1)}(x,s)$, and has the general form
\begin{equation}
\psi _{\gamma }^{(2)}(x_{1},x_{2},s)=\int dx\,K(x_{1},x_{2},s;x)\psi
_{\gamma }^{(1)}(x,s).  \label{3.9}
\end{equation}%
The kernel defining the functional is
\begin{equation}
K(x_{1},x_{2},s;x)=\left[ \frac{\delta f^{(2)}(x_{1},x_{2},s)}{\delta
f^{(1)}(x,s)}\right] _{\delta y=0}.  \label{3.10}
\end{equation}%
Once $K(x_{1},x_{2},s;x)$ is known, Eq.\ (\ref{3.5}) becomes a closed,
deterministic, linear kinetic equation for $\psi _{\gamma }^{(1)}$ in its
most general form,
\begin{equation}
\left[ \frac{\partial }{\partial s}+{\bm v}_{1}\cdot \frac{\partial }{%
\partial {\bm q}_{1}}+M(s)\right] \psi _{\gamma }^{(1)}(x_{1},s)+\frac{\zeta
_{0}}{2}\frac{\partial }{\partial {\bm v}_{1}}\cdot \left[ {\bm v}_{1}\psi
_{\gamma }^{(1)}(x_{1},s)\right] =0,  \label{3.11}
\end{equation}%
with the formal \textquotedblleft collision operator\textquotedblright\ $M$
given by
\begin{equation}
M\left( x_{1},s\right) \psi _{\gamma }^{(1)}(x_{1},s)\equiv -\int dx\,\left[
\int dx_{2}\,\overline{T}(x_{1},x_{2})K(x_{1},x_{2},s;x)\right] \psi
_{\gamma }^{(1)}(x,s).  \label{3.12}
\end{equation}%
This notion of a kinetic equation makes no \emph{a priori} assumptions
regarding the density or degree of dissipation, as is sometimes assumed. In
fact, it is formally exact so any restrictions arise only when specific
approximations are introduced to construct the functional.

The utility of the kinetic theory representation in any specific application
depends on an accurate construction of the kernel $K(x_{1},x_{2},s;x)$
defined in Eq.\ (\ref{3.10}). This is the point at which the full many-body
problem must be confronted. The idea originated with Bogoliubov \cite%
{Bogoliubov46}, who conceived that all $f^{(m)}(x_{1},\ldots,x_{m},s)$ with $%
m>1$ take a simpler form, $f^{(m)}[x_{1},..,x_{m} | f^{(1)}(s)]$, after a
brief ``synchronization'' time, where all time dependence occurs only
through $f^{(1)}(s)$. For normal fluids at low density, construction of this
functional can be accomplished by formal density expansions, leading to a
sequence of contributions from clusters of particles of increasing size \cite%
{Cohen}. At lowest order, the closure is that associated with the Boltzmann
equation, while at next order three particle scattering is described. Even
for normal fluids, these sophisticated cluster expansions have limited
direct use due to many-particle recollisions events that contribute secular
terms in the formal expansions, violating the notion of a short
synchronization time \cite{recollisions}. These recollisions (``rings'') in
turn signal non-analytic density dependence and slow algebraic decay of
correlations in time \cite{Alder2}. The experience gained in such studies
over the past forty years has provided important insight for the
construction of more phenomenological closures, such as the Enskog
approximation for moderately dense gases and mode coupling models for very
dense and metastable (glassy) fluids \cite{Modecoupling}. The development of
a kinetic theory for granular fluids provides an opportunity to revisit many
of these issues in an even more challenging context \cite{vanNoije01}. The
analysis of the next sections illustrates this for the simplest ``mean
field'' approximation to describe density effects beyond the Boltzmann limit.

\section{The Markovian Approximation}

\label{s4}

The approximation described in this section is based on the assumption that
the form of the correlations and their effect on collisional properties are
essentially the same at all times. If so, the kernel $K(x_{1},x_{2},s;x) $
that determines these properties at the two-particle level, can be
represented approximately by its form at the initial time $s=0$,
\begin{equation}
K(x_{1},x_{2},s;x)\simeq K(x_{1},x_{2},0;x).  \label{4.1}
\end{equation}
The collision operator term in Eq.\ (\ref{3.12}) then becomes
\begin{equation}
M(x_{1}) \psi ^{(1)}_{\gamma}\left( x_{1},s\right) = -\int dx\, \int
dx_{2}\, \overline{ T}(x_{1},x_{2}) \left[ \frac{\delta f_{\ell h}^{(2)}%
\left[ x_{1},x_{2}|\delta y \right]}{\delta f_{\ell h }^{(1)}[x|\delta y]} %
\right]_{\delta y=0} \psi ^{(1)}_{\gamma}(x,s),  \label{4.6}
\end{equation}
where it has been used that the initial conditions $%
f^{(m)}(x_{1},..,x_{m},0) $ are the corresponding reduced distribution
functions associated with the local HCS, $\rho _{\ell h }\left[ \Gamma
|\delta y\right] $, given in Eq.\ (\ref{a.7}),
\begin{equation}
f^{(m)}(x_{1},..,x_{m},0)=f_{\ell h}^{(m)}[x_{1},..,x_{m}| \delta y ])=
\frac{N!}{(N-m)!}\int dx_{m+1}..dx_{N}\, \rho _{\ell h}\left[ \Gamma |\delta
y\right].
\end{equation}
Since the approximate operator $M(x_{1})$ is time-independent, the entire
generator for the dynamics of $\psi ^{(1)}_{\gamma}(x_{1},s)$ in Eq.\ (\ref%
{3.11}) also is independent of time. The equation is written in compact form
as
\begin{equation}
\left[ \frac{\partial}{\partial s}+\Lambda (x_{1}) \right] \psi
^{(1)}_{\gamma}(x_{1},s)=0,  \label{4.5}
\end{equation}
with the generator $\Lambda(x_{1})$ given by
\begin{equation}
\Lambda(x_{1}) X(x_{1}) \equiv \left[ {\bm v}_{1}\cdot \frac{\partial}{%
\partial {\bm q}_{1}}+M (x_{1})\right] X (x_{1})+\frac{\zeta_{0}}{2} \frac{%
\partial}{\partial {\bm v}_{1}} \cdot \left[ {\bm v}_{1}X(x_{1})\right],
\label{4.5a}
\end{equation}
for arbitrary $X(x_{1})$. This is a necessary condition for a Markovian
description and, consequently, Eq.\ (\ref{4.1}) will be referred to as the
Markovian approximation. The resulting kinetic theory is exact at
asymptotically short times, and the nature of the approximation makes no
explicit limitation on the density or the degree of dissipation. Of course,
two particle correlations that develop over time are neglected. For a normal
fluid, this Markov approximation leads to the Enskog approximation, where
only time independent two-particle correlations are taken into account. The
neglected time-dependent correlations are found in that case to be important
only at high densities, and the Enskog approximation provides relevant
corrections to the Boltzmann results up to moderate densities. It is
reasonable to expect a similar context for granular fluids, although
conditioned by the additional parameter space of the coefficient of
restitution $\alpha $.

The construction of $K(x_{1},x_{2},0;x)$ now involves only analysis of the
initial reduced distribution functions associated with the local HCS, and
the corresponding collision operator $M$ is entirely characterized by
properties of this state. It is useful to make this more explicit in terms
of the pair correlation function $g_{\ell h }^{(2)}[x_{1},x_{2}|\delta y]$
for the HCS, defined by
\begin{equation}
f_{\ell h}^{(2)}[x_{1},x_{2}|\delta y]\equiv f_{\ell h}^{(1)}[x_{1}|\delta
y]\, f_{\ell h}^{(1)}[x_{2}|\delta y]\, g_{\ell h }^{(2)}[x_{1},x_{2}|\delta
y].  \label{4.7}
\end{equation}
Then Eq.\ (\ref{3.10}) in the Markovian approximation yields
\begin{eqnarray}
K(x_{1},x_{2},0;x)& = & g_{h}^{(2)}(x_{1},x_{2})\left[ f_{h}^{(1)}(v_{1})%
\delta \left( x-x_{2}\right) + f_{h}^{(1)}(v_{2}) \delta \left(
x-x_{1}\right) \right]  \notag \\
& & + f_{h}^{(1)}(v_{1})f_{h}^{(1)}(v_{2}) \left[ \frac{\delta g_{\ell h
}^{(2)}[x_{1},x_{2}|\delta y]}{\delta f_{\ell h }^{(1)}[x|\delta y]} \right]%
_{\delta y=0}\, .  \label{4.8}
\end{eqnarray}
The first term on the right side is given explicitly in terms of the HCS
correlations, while the second term requires further analysis of the
dependence of local HCS correlations on $f_{\ell h}^{(1)}$. For a \emph{%
normal fluid}, these correlations are independent of the velocities and
depend on the local equilibrium distribution $f_{\ell e }^{(1)}$ only
through the local density,
\begin{equation}
\left[ \frac{\delta g_{\ell h}^{(2)}\left[ x_{1},x_{2}|\delta y\right] }{%
\delta f_{\ell h}^{(1)}[x|\delta y]} \right] _{\delta y=0} \rightarrow \frac{%
1}{n_{h}}\left[ \frac{\delta g_{\ell e }^{(2)}[{\bm q}_{1},{\bm q}%
_{2}|\delta y]}{\delta n({\bm q}) }\right]_{\delta y=0}\, .  \label{4.9}
\end{equation}
The functional form of the local equilibrium pair correlation functional $%
g_{\ell e }^{(2)}[{\bm q}_{1},{\bm q}_{2}|\delta y ]$ is known, so explicit
construction of the collision operator $M$ is possible in this case. The
result is the linear revised Enskog kinetic equation \cite{vanBeijeren73}.
For granular fluids, $g_{\ell h}^{(2)}[x_{1},x_{2}|\delta y]$ has a more
general functional functional dependence on $f_{\ell h }^{(1)}$ through its
explicit functional dependence on $\delta y,$ since
\begin{equation}
\delta y_{\gamma }\left({\bm q}\right) =\frac{1}{n_{h}}\int dx_{1}\,
a_{\gamma }\left( x_{1}\right) \delta \left( {\bm q}-{\bm q}_{1}\right)
\left\{ f_{\ell h}^{(1)}[x_{1}|\delta y]-f_{h}^{(1)}(v_{1}) \right\},
\label{4.9a}
\end{equation}
where the $a_{\gamma }\left( x_{1}\right) $'s are the single particle
functions given in Eq.\ (\ref{2.23}). Thus
\begin{equation}
\left[ \frac{\delta g_{\ell h}^{(2)}[x_{1},x_{2}|\delta y]}{\delta f_{\ell
}^{(1)}[x|\delta y]}\right]_{\delta y=0}=\frac{1}{n_{h}} \sum_{\lambda} %
\left[ \frac{\delta g_{\ell h }^{(2)}[x_{1},x_{2}|\delta y]}{\delta
y_{\lambda }({\bm q})}\right] _{\delta y=0}a_{\lambda }\left({\bm v} \right)
.  \label{4.9b}
\end{equation}
This provides the practical route for constructing the Markovian kinetic
theory for granular fluids. The collision operator becomes
\begin{eqnarray}
M(x_{1}) \psi _{\gamma }^{(1)}(x_{1},s) &= &-\int dx_{2}\, \overline{T}
(x_{1},x_{2})g_{h}^{(2)}(x_{1},x_{2})\left[ f_{h}^{(1)}(v_{1})\psi _{\gamma
}^{(1)}(x_{2},s)+f_{h}^{(1)}(v_{2}) \psi _{\gamma }^{(1)}(x_{1},s)\right]
\notag \\
&&- \sum_{\lambda} \int d{\bm q}_{2}\, c_{\lambda }({\bm v}_{1,}{\bm q}_{12})%
\frac{1}{n_{h}} \int d{\bm v}_{2}\, a_{\lambda }\left( {\bm v}_{2}\right)
\psi _{\gamma }^{(1)}(x_{2},s),  \label{4.10}
\end{eqnarray}
with
\begin{equation}
c_{\lambda }({\bm v}_{1},{\bm q}_{12})=\int dx\overline{T }%
(x_{1},x)f_{h}^{(1)}(v_{1})f_{h}^{(1)}(v)\left[ \frac{\delta g_{\ell
h}^{(2)}[x_{1},x|\delta y]}{\delta y_{\lambda }({\bm q}_{2})}\right]
_{\delta y=0}.  \label{4.11}
\end{equation}
Further discussion and simplification of $c_{\lambda }({\bm
v}_{1,} {\bm q}_{12})$ is given in Appendix \ref{ap3}.

The response functions of Eq.\ (\ref{3.1}) are given in terms of $\widetilde{%
\psi }_{\gamma }^{(1)}({\bm v}_{1},-{\bm k},s)$, that is proportional to the
Fourier transform of $\psi _{\gamma }^{(1)}(x_{1},s)$ as indicated in Eq.\ (%
\ref{3.3}). An equation for the latter can be easily derived from Eq.\ (\ref%
{3.11}),
\begin{equation}
\left[ \frac{\partial }{\partial s}+\widetilde{\Lambda }({\bm k})\right]
\widetilde{\psi }_{\gamma }^{(1)}({\bm v}_{1},-{\bm k},s)=0,  \label{4.15}
\end{equation}%
where the generator for the dynamics is the Fourier transform of $\Lambda $
defined in Eq.\thinspace\ (\ref{4.5a}),
\begin{equation}
\widetilde{\Lambda }({\bm k})=-i{\bm k}\cdot {\bm v}_{1}+\widetilde{M}\left(
{\bm k}\right) +\frac{\zeta _{0}}{2}\left( d+{\bm v}\cdot \frac{\partial }{%
\partial {\bm v}}\right) .  \label{4.16}
\end{equation}%
with $\widetilde{M}({\bm v}_{1},{\bm k})$ defined by
\begin{eqnarray}
\widetilde{M}\left( {\bm v}_{1},{\bm k}\right) X({\bm v}_{1}) &\equiv &-\int
dx_{2}\,\overline{T}(x_{1},x_{2})g_{h}^{(2)}(x_{1},x_{2})\left[ e^{i{\bm k}%
\cdot {\bm q}_{12}}f_{h}^{(1)}(v_{1})X({\bm v}_{2})+f_{h}^{(1)}(v_{2})X({\bm %
v}_{1})\right]  \notag \\
&&-\sum_{\lambda }\widetilde{c}_{\lambda }({\bm v}_{1,}{\bm k})\frac{1}{n_{h}%
}\int d{\bm v}_{2}\,a_{\lambda }\left( {\bm v}_{2}\right) X({\bm v}_{2}),
\label{4.14}
\end{eqnarray}%
for arbitrary $X({\bm v}_{1})$, $\widetilde{c}_{\lambda }({\bm v}_{1,}{\bm k}%
)$ being the Fourier transform of $c_{\lambda }({\bm v}_{1},{\bm q}_{12})$.

The response functions in Eq.\ (\ref{3.1}) become
\begin{equation}
\widetilde{C}_{\beta \gamma }\left( {\bm k},s\right) =\frac{1}{n_{h}}\int d{%
\bm v}_{1}a_{\beta }\left( {\bm v}_{1}\right) e^{-s\widetilde{\Lambda }({\bm %
k})}\widetilde{\psi }_{\gamma }^{(1)}({\bm v}_{1},{\bm k}),  \label{4.17}
\end{equation}%
\begin{equation}
\widetilde{\psi }_{\gamma }^{(1)}({\bm v}_{1},-{\bm k})=\int d{\bm q}_{1}e^{i%
{\bm k}\cdot {\bm q}_{1}}\,\left[ \frac{\delta f_{\ell h}^{(1)}[x_{1}|\delta
y]}{\delta \widetilde{y}_{\gamma }({\bm0})}\right] _{\delta \widetilde{y}=0}.
\label{4.17a}
\end{equation}

Equation (\ref{4.17}) is the primary practical result of our analysis here.
It provides a realistic kinetic theory description of the most fundamental
time dependent fluctuations in a granular fluid: those induced by
perturbations of the hydrodynamic fields. To appreciate the scope and
generality of this result, note that the Markovian approximation is exact at
short times for all ${\bm k}$, densities, and degrees of restitution. In
this short time limit, Eq.\ (\ref{4.17}) yields
\begin{equation}
\lim_{s\rightarrow 0}\widetilde{C}_{\beta \gamma }\left( {\bm k},s\right)
=\lim_{s\rightarrow 0}\left( e^{-sN\left( {\bm k}\right) }\right) _{\beta
\gamma },  \label{4.18}
\end{equation}%
with
\begin{equation}
N_{\beta \gamma }\left( {\bm k}\right) =\frac{1}{n_{h}}\int d{\bm v}%
_{1}a_{\beta }\left( {\bm v}_{1}\right) \widetilde{\Lambda }({\bm k})%
\widetilde{\psi }_{\gamma }^{(1)}({\bm v}_{1},{\bm k}).  \label{4.19}
\end{equation}%
At longer times, the Markovian approximation is expected to continue to
provide a good approximation across this parameter space, since the
derivation does not explicitly require any limitations on ${\bm k}$,
densities, or degrees of restitution. This expectation is borne out in the
elastic limit, where the linear Enskog theory for equilibrium time
correlation functions is recovered, as discussed in the next section. In
that case, comparisons with molecular dynamics simulations (wavevector
dependent transport \cite{Alley}) and neutron scattering experiments \cite%
{deSchepper}, confirm the accuracy and practical utility of this kinetic
theory over a wide range of wavevectors and densities. A determination of
the corresponding domain of accuracy for granular fluids, awaits similar
comparisons with simulation and experiments, but there is no simple reason
to expect qualitative rather than quantitative differences from normal
fluids.

In Sec. \ref{s6}, it is shown that the operator $\widetilde{\Lambda }({\bm k}
) $ contains the hydrodynamic modes in its spectrum for small $k$. Thus the
response functions in Eq.\ (\ref{4.17}) provide a means to study the
transition from short time dynamics to a presumed dominant hydrodynamics at
long times for granular fluids, in a manner similar to that done for normal
fluids \cite{McL89}. In addition, since it is valid for all ${\bm
k}$, the nature of hydrodynamics beyond the Navier-Stokes approximation can
be studied.

An important application of general linear response methods, is the
derivation of Helfand and Green-Kubo expressions for the transport
coefficients \cite{DBB06,BDB06}. These are formally exact results given in
terms of time correlation functions. The above analysis for the response
functions applies to these as well, and their dynamics in the Markovian
approximation is generated by the same operator $\widetilde{\Lambda }({\bm k}%
)$. The evaluation of the Helfand and Green-Kubo expressions for the shear
viscosity is illustrated in Sec. \ref{s7}. The results provide a
generalization of those from Enskog kinetic theory \cite{GD99}, to include
pair velocity correlations. When such correlations are neglected, the
results of ref. \cite{GD99} are recovered in detail. This is verified for
the other Navier-Stokes transport coefficients as well in Appendix \ref{ap5}.

\section{Granular Enskog Approximation}

\label{s5}

The Markovian approximation discussed in the previous section, requires
specification of $g_{ \ell h}^{(2)}(x_{1},x_{2})$ and $c_{\lambda }(x_{1},{%
\bm q}_{12})$, or equivalently $\delta g_{\ell h}^{(2)}[x_{1},x|\delta
y]/\delta y_{\lambda }( {\bm
q}_{2})$ for $\delta y=0$. While these are well defined in terms of the
local HCS distribution, little is know about their detailed forms as yet,
except in the elastic limit where they are accurately determined from liquid
state theory of the pair correlation function. The important simplification
in that case is the absence of velocity correlations. It is plausible to
assume that such correlations remain weak for the granular fluid as well,
and to make the approximation
\begin{equation}
g_{\ell h }^{(2)}[x_{1},x_{2}|\delta y]\simeq g_{\ell h }^{(2)}[{\bm q}_{1},
{\bm q}_{2}|\delta y]=g_{\ell h}^{(2)}[{\bm q}_{1},{\bm q}_{2}|\delta n].
\label{5.1}
\end{equation}
The last equality recognizes that the neglect of velocity correlations leads
to a functional that is independent of $\delta T$ and $\delta {\bm U}$ (for
hard spheres or disks) and hence is a functional only of the density.
Furthermore, $g_{\ell h }^{(2)}[{\bm q}_{1},{\bm q}%
_{2}|0]=g_{h}^{(2)}(q_{12})$ as a consequence of fluid symmetry. Then, Eq.\ (%
\ref{4.11}) reduces to
\begin{equation}
c_{\lambda }^{(E)}({\bm v}_{1,}{\bm q}_{12})= \delta _{\lambda 1}c^{(E)}( {%
\bm v}_{1, },{\bm q}_{12}) = \delta _{\lambda 1}\int dx\, \overline{T}
(x_{1},x)f_{h}^{(1)}(v_{1})f_{h}^{(1)}(v) \left[ \frac{\delta g_{\ell
h}^{(2)}( {\bm q}_{1},{\bm q}|\delta n)}{\delta n({\bm q}_{2})}\right]
_{\delta n=0}.  \label{5.2}
\end{equation}
The collision operator (\ref{4.14}) now simplifies to
\begin{eqnarray}
\widetilde{M}^{E} X({\bm v}_{1}) & = &-g_{h}^{(2)}(\sigma )\int dx_{2}\,
\overline{T}(x_{1},x_{2})\left[ e^{- i {\bm k} \cdot {\bm q}_{12}
}f_{h}^{(1)}(v_{1}) X({\bm v}_{2})+f_{h}^{(1)}(v_{2}) X({\bm v}_{1})\right]
\notag \\
&&- \frac{1}{n_{h}} \widetilde{c}^{(E)}({\bm v}_{1},{\bm k})\int d{\bm v}%
_{2}\, X({\bm v}_{2}).  \label{5.3}
\end{eqnarray}
It remains to give the explicit density dependence for $g_{\ell h }^{(2)}[ {%
\bm r}_{1},{\bm r}|\delta n]$ \cite{lutsko01}. As a practical matter, it can
be chosen to be the pair distribution for a nonuniform normal fluid for
which a well-developed theory exists. In that case, $g_{h}^{(2)}(\left| {\bm %
r}_{1}-{\bm r}\right| ) = g_{eq}^{(2)}(\left| {\bm r} _{1}-{\bm r}\right| )$%
, the radial distribution function for a uniform hard sphere fluid. Also,
for this choice the functional derivative appearing in the expression of $%
\widetilde{c}^{(E)}({\bm v}_{1},{\bm k})$ can be evaluated for the first few
terms of a ${\bm k}$ expansion, as is required for evaluation of transport
coefficients. The context of such a choice would be that the static spatial
correlations of a hard sphere system are due to excluded volume effects, and
these can be captured using the pair correlation function of a fluid of
elastic hard spheres. The generator $\Lambda({\bm k})$ obtained in this
approximation from Eq.\ (\ref{5.3}) gives the granular Enskog kinetic
theory, and is the linearized version of the one studied in ref. \cite{GD99}.

\section{Hydrodynamic Modes}

\label{s6}

An important feature of the response functions considered here is their
relationship to hydrodynamic response. At small $k$ and large $s$, these
response functions should correspond to those from the phenomenological
hydrodynamic equations. For example, it is this relationship that allows the
derivation of Helfand and Green-Kubo expressions for the transport
coefficients. Any acceptable approximate kinetic theory for the response
functions should preserve this relationship to hydrodynamics. More
specifically, the hydrodynamic excitations should appear in the spectrum of
the linear operator $\widetilde{ \Lambda }({\bm k})$,
\begin{equation}
\widetilde{\Lambda }({\bm k},{\bm v})\phi ^{(\beta)}({\bm k},{\bm v}%
)=\lambda ^{(\beta)}( {\bm k})\phi ^{(\beta)}({\bm k},{\bm v}), \quad \beta
= 1,\ldots, d+2,  \label{6.1}
\end{equation}
where the set $\left\{ \lambda ^{(\beta)}({\bm k})\right\} $ are the
eigenvalues of the $d+2$ linearized hydrodynamic equations. The above
eigenfunctions and eigenvalues are determined in the limit ${\bm k}={\bm 0}$
in Appendix \ref{ap4} with the results
\begin{equation}
\left\{ \lambda ^{(\beta)}({\bm 0}) \right\} = \left\{ 0,\frac{\zeta_{0}}{2}%
,-\frac{\zeta_{0}}{2} \right\} ,  \label{6.2}
\end{equation}
\begin{equation}
\left\{ \phi^{(\beta)} ({\bm v},{\bm 0}) \right\} = \left\{ \widetilde{\psi}%
_{1}^{(1)} ({\bm v},{\bm 0})-2 \left( \frac{\partial \ln \zeta_{0}}{\ln n_{h}%
} \right)_{T_{h}} \widetilde{\psi}_{2}^{(1)} ({\bm v},{\bm 0}),\widetilde{%
\psi}_{2}^{(1)} ({\bm v},{\bm 0}),\widetilde{\bm \psi}_{3}^{(1)} ({\bm v},{%
\bm 0}) \right\}.  \label{6.3}
\end{equation}
The eigenvalue $-\zeta_{0}/2$ is $d$-fold degenerate, and the associated
eigenfunctions are the components of the vector $\widetilde{\bm \psi}%
_{3}^{(1)} ({\bm v},{\bm 0}) \equiv - \partial f_{h}^{(1)}(v) / \partial {%
\bm v}$.

In the elastic limit, these eigenvalues are all zero, corresponding to the $%
d+2$ conservation laws, and the eigenfunctions become Maxwellians times
linear combinations of the summational invariants $\left( 1,v^{2},{\bm v}
\right) $. For inelastic collisions, the nonzero eigenvalues describe
response of the cooling temperature to linear perturbations and growth of a
constant velocity perturbation relative to the characteristic cooling
thermal velocity. In both cases, these are also the eigenvalues of the
phenomenological linearized hydrodynamic equations in the long wavelength
limit.

With the eigenfunctions and eigenvalues known at ${\bm k}=0$, their values
for finite but small $k$ can be obtained by perturbation theory. In this way
the Navier-Stokes transport coefficients can be determined directly from the
coefficients up through order $k^{2}$. This direct calculation of the
spectrum for the generator of a linear kinetic theory has been described in
detail recently for the granular Boltzmann equation \cite{Dufty03}, and its
extension to the Markovian kinetic theory given here is straightforward.
Instead, the remainder of this presentation addresses the calculation of the
transport coefficients from an approximate evaluation of their Helfand and
Green-Kubo representations that have been obtained from linear response.

\section{Helfand and Green-Kubo Expressions}

\label{s7}

As noted above, the exact response functions defined in Sec. \ref{s2} must
agree with those of the linearized phenomenological hydrodynamic equations
in the long wavelength and long time limit. This relationship allows
identification of the parameters of those phenomenological equations in
terms of the response functions in this limit. The results of this analysis
for granular fluids has been given recently, leading to expressions for the
transport coefficients in terms of certain time correlation functions
derived from the response functions \cite{BDB06}. These correlation
functions can be evaluated approximately by the Markov kinetic theory, to
obtain explicit results for all transport coefficients appearing in the
Navier-Stokes hydrodynamic equations. Further, neglecting the velocity
correlations in the Markov theory, allows the evaluation of these quantities
in the granular Enskog theory, reproducing the results reported in \cite%
{GD99}. In this section, only the shear viscosity is considered as an
example, while all remaining transport coefficients are analyzed in Appendix %
\ref{ap5}.

The exact Helfand and Green-Kubo expressions for the shear viscosity are
(dimensionless units are still assumed)
\begin{equation}
\eta =\lim \Omega _{H}^{\eta }\left( s\right) =\Omega _{H}^{\eta }\left(
0\right)+\lim \int_{0}^{s}ds^{\prime }\, \Omega _{G}^{\eta }\left( s^{\prime
}\right),  \label{7.1}
\end{equation}
respectively, where the symbol $\lim $ denotes the hydrodynamic limit of $%
V\rightarrow \infty $, followed by $s \rightarrow \infty$. The correlation
functions in the above equation are defined by
\begin{equation}
\Omega _{H}^{\eta }\left( s\right) =- \frac{V^{-1}}{d^{2}+d-2}
\sum_{i=1}^{d} \sum_{j=1}^{d} \int d\Gamma\, H_{ij}(\Gamma) e^{-s \left(
\overline{\mathcal{L}}+\frac{\zeta _{0}}{2}\right) } \mathcal{M}_{\eta
,ij}(\Gamma),  \label{7.2}
\end{equation}
\begin{equation}
\Omega _{G}^{\eta }\left( s\right) =\frac{\partial}{\partial s} \Omega
_{H}^{\eta }\left( s\right)=- \frac{V^{-1}}{d^{2}+d-2} \sum_{i=1}^{d}
\sum_{j=1}^{d} \int d\Gamma\, H_{ij}(\Gamma) e^{-s \left( \overline{\mathcal{%
L}}+\frac{\zeta _{0}}{2}\right) }\Upsilon _{\eta ,ij}(\Gamma).  \label{7.3}
\end{equation}
Here, $H_{ij}(\Gamma)$ is the volume integrated momentum flux,
\begin{equation}
H_{ij}(\Gamma) =\sum_{r=1}^{N}v_{r,i}v_{r,j}+ \sum_{r=1}^{N} \sum_{s \neq
r}^{N}H_{ij}^{(2)}\left( x_{r},x_{r}\right) ,  \label{7.3a}
\end{equation}
\begin{equation}
H_{ij}^{(2)}\left( x_{r},x_{s}\right) =\frac{(1+\alpha)\sigma}{4} \delta
\left( q_{rs}-\sigma \right) \Theta \left( -\widehat{\bm q} _{rs}\cdot {\bm g%
}_{rs}\right) \left( \widehat{\bm q}_{rs}\cdot {\bm g}_{lm}\right) ^{2}%
\widehat{q}_{rs,i}\widehat{q}_{rs,j},  \label{7.3b}
\end{equation}
$\mathcal{M}_{\eta ,ij}\mathcal{\ }$is the traceless tensor
\begin{equation}
\mathcal{M}_{\eta ,ij}=- \frac{1}{2}\sum_{r=1}^{N} \left( q_{ri} \frac{%
\partial}{\partial v_{r,j}}+ q_{r,j} \frac{\partial}{\partial v_{r,i}} -
\frac{2}{d} \delta_{ij} {\bm q}_{r} \cdot \frac{\partial}{\partial {\bm v}%
_{r}} \right) \rho_{h}(\Gamma),  \label{7.4}
\end{equation}
and $\Upsilon _{\eta ,ij}(\Gamma)$ is the associated Green-Kubo conjugate
flux,
\begin{equation}
\Upsilon _{\eta ,ij}(\Gamma) =-\left( \overline{\mathcal{L}}+\frac{\zeta _{0}%
}{2} \right) \mathcal{M}_{\eta ,ij}(\Gamma).  \label{7.5}
\end{equation}
It is seen in Eqs.\ (\ref{7.2}) and (\ref{7.3}) that the generator of
dynamics is $\overline{\mathcal{L}}+\frac{\zeta _{0} }{2} $. This reflects
the fact that the $k=0$ mode $\lambda =-\zeta _{0}/2,$ of Eq.\ (\ref{6.2}),
has been subtracted out.

The time independent contribution\ $\Omega _{H}^{\eta }\left( 0\right) $ in
the Green-Kubo expression, can be evaluated exactly from the definitions (%
\ref{7.3a}) and (\ref{7.4}) with the result:
\begin{equation}
\Omega _{H}^{\eta }\left( 0\right)=\frac{V^{-1}}{d^{2}+d-2} \sum_{i=1}^{d}
\sum_{j=1}^{d} \int dx_{1}\, h_{ij}({\bm v}_{1}) \frac{1}{2} \left( q_{1,i}
\frac{\partial}{\partial v_{1,j}} +q_{1,j} \frac{\partial}{\partial v_{1,y}}
-\frac{2}{d} \delta_{ij} {\bm q}_{1} \cdot \frac{\partial}{\partial {\bm v}%
_{1}} \right) f_{h}^{(1)}(v_{1}),  \label{7.6}
\end{equation}
where
\begin{equation}
h_{ij}({\bm v})=v_{i}v_{j}+\int dx_{1} \int dx_{2}\, H_{ij}^{(2)}\left(
x_{1},x_{2}\right) K(x_{1},x_{2},0;x).  \label{7.6a}
\end{equation}
The first term in the above expression of $h_{ij}$ gives no contribution to $%
\Omega_{H}^{\eta}(0)$, from fluid symmetry. The second term can be
recognized as being proportional to the average collision frequency, $\nu
_{av}$, as determined by the loss part of the right hand side of the hard
sphere BBGKY hierarchy (\ref{3.3c}) specialized for the HCS,
\begin{equation}
\Omega _{H}^{\eta }\left( 0\right)=\frac{(1+\alpha)\sigma^{2}}{4(d^{2}+2d)}%
\, \nu _{av}  \label{7.7}
\end{equation}
\begin{equation}
\nu _{av}=2 \sigma ^{d-1}\int d\widehat{\bm \sigma }\int d{\bm v}_{1} \int d
{\bm v}_{2}\, \Theta \left( -\widehat{\bm \sigma }\cdot {\bm g }_{12}\right)
\left( \widehat{\bm \sigma }\cdot {\bm g}_{12}\right) f_{h}^{\left( 2\right)
}\left({\bm \sigma },{\bm v}_{1}, {\bm v}_{2}\right) .  \label{7.8}
\end{equation}

\subsection{Evaluation in the Markov Approximation}

A complete evaluation of the correlation function $\Omega _{H}^{\eta }\left(
s\right) $ is possible using the Markov kinetic theory. As in Sec. \ref{s3},
the correlation function can be given a representation in terms of one and
two particle functions
\begin{eqnarray}
\Omega _{H}^{\eta }\left( s\right) & = & -\frac{V^{-1}}{d^{2}+d-2}
\sum_{i=1}^{d} \sum_{j=1}^{d} \int dx_{1}\, v_{1,i}v_{1,j} \mathcal{M}_{\eta
,ij}^{(1)}\left( x_{1},s\right)  \notag \\
& - & \frac{V^{-1}}{d^{2}+d-2} \sum_{i=1}^{d} \sum_{j=1}^{d} \int dx_{1}
\int dx_{2}\, H_{ij}^{(2)}\left( x_{1},x_{2}\right) \mathcal{M}_{\eta
,ij}^{(2)}\left( x_{1},x_{2},s\right) ,  \label{7.9}
\end{eqnarray}
with the reduced functions $\mathcal{M}_{\eta ,ij}^{(m)}\left(
x_{1},..,x_{m},s\right) $ defined by
\begin{equation}
\mathcal{M}_{\eta ,ij}^{(m)}\left( x_{1},\ldots ,x_{m},s\right) \equiv \frac{%
N!}{ (N-m)!}\int dx_{m+1}\ldots \int dx_{N}e^{-s \left( \overline{\mathcal{L}%
}+\frac{\zeta _{0}}{2}\right) }\mathcal{M}_{\eta ,ij}(\Gamma).  \label{7.10}
\end{equation}
Similarly to the functions $\psi_{\gamma}$ considered in Sec.\ \ref{s3}, the
above functions obey a BBGKY hierarchy, the first equation of which is
\begin{eqnarray}
\left( \frac{\partial}{\partial s} + \frac{\zeta _{0}}{2}+ {\bm v}_{1}\cdot
\frac{\partial}{\partial {\bm q}_{1}}\right) \mathcal{M}_{\eta
,ij}^{(1)}(x_{1},s)& + & \frac{\zeta_{0}}{2} \frac{\partial}{\partial{\bm v}%
_{1}} \cdot \left[ {\bm v}_{1} \mathcal{M}_{\eta ,ij}^{(1)}(x_{1},s)\right]
\notag \\
& = & \int dx_{2}\, \overline{T}(x_{1},x_{2})\mathcal{M}_{\eta
,ij}^{(2)}(x_{1},x_{2},s).  \label{7.11}
\end{eqnarray}
The Markovian approximation in the present case is the same as that defined
by Eqs.\ (\ref{3.9}) and (\ref{4.1})
\begin{equation}
\mathcal{M}_{\eta ,ij}^{(2)}(x_{1},x_{2},s) \simeq \int dx\,
K(x_{1},x_{2},0;x)\mathcal{M}_{\eta ,ij}^{(1)}(x,s).  \label{7.12}
\end{equation}
Then, Eq.\ (\ref{7.11}) becomes the Markovian kinetic equation
\begin{equation}
\left( \frac{\partial}{\partial s}+\frac{\zeta _{0}}{2}+\Lambda \right)
\mathcal{M}_{\eta ,ij}^{(1)}=0,  \label{7.13}
\end{equation}
where the linear operator $\Lambda $ is the same as defined in Eq.\ (\ref%
{4.5a}).

In this approximation, the Helfand expression of the shear viscosity of a
hard sphere or disk granular fluid becomes
\begin{eqnarray}
\eta = \lim \Omega _{H}^{\eta }\left( s\right) & \simeq & \frac{V^{-1}}{%
d^{2}+d-2} \sum_{i=1}^{d} \sum_{j=1}^{d}\int dx_{1}\, h_{ij}({\bm v}%
_{1})e^{-s \left( \Lambda +\frac{\zeta _{0}}{2}\right) }  \notag \\
&& \times \frac{1}{2} \left( q_{1,i} \frac{\partial}{\partial v_{1,j}}
+q_{1,j} \frac{\partial}{\partial v_{1,y}} -\frac{2}{d} \delta_{ij} {\bm q}%
_{1} \cdot \frac{\partial}{\partial {\bm v}_{1}} \right) f_{h}^{(1)}(v_{1}),
\label{7.16a}
\end{eqnarray}
where $h_{ij}({\bm v})$ is defined in Eq.\ (\ref{7.6a}).

Next, the Green-Kubo expression for $\eta $ in the Markov approximation can
be identified from Eqs.\ (\ref{7.1}) and (\ref{7.3}),
\begin{equation}
\Omega _{H}^{\eta }\left( 0\right) = \frac{V^{-1}}{d^{2}+d+2} \sum_{i=1}^{d}
\sum_{j=1}^{d} \int dx_{1\, }h_{ij}({\bm v}_{1}) \frac{1}{2} \left( q_{1,i}
\frac{\partial}{\partial v_{1,j}} +q_{1,j} \frac{\partial}{\partial v_{1,y}}
-\frac{2}{d} \delta_{ij} {\bm q}_{1} \cdot \frac{\partial}{\partial {\bm v}%
_{1}} \right) f_{h}^{(1)}(v_{1})  \label{7.17a}
\end{equation}
and
\begin{equation}
\Omega _{G}^{\eta }\left( s\right) = \frac{V^{-1}}{d^{2}+d-2} \sum_{i=1}^{d}
\sum_{j=1}^{d} \int dx_{1}\, h_{ij}({\bm v}_{1})e^{-s \left( \Lambda +\frac{%
\zeta _{0}}{2}\right) }\gamma _{ij}\left( {\bm v}_{1}\right).  \label{7.18}
\end{equation}
The reduced conjugate flux $\gamma _{ij}$ is
\begin{equation}
\gamma _{ij}\left( {\bm v}_{1}\right) =-\left( \Lambda +\frac{\zeta _{0}}{ 2}%
\right) \frac{1}{2} \left( q_{1,i} \frac{\partial}{\partial v_{1,j}}
+q_{1,j} \frac{\partial}{\partial v_{1,y}} -\frac{2}{d} \delta_{ij} {\bm q}%
_{1} \cdot \frac{\partial}{\partial {\bm v}_{1}} \right) f_{h}^{(1)}(v_{1}).
\label{7.19a}
\end{equation}
Comparison of Eqs.\ (\ref{7.6}) and (\ref{7.17a}) shows that $\Omega
_{H}^{\eta }\left( 0\right) $ is given exactly in the Markov approximation.
The Green-Kubo representation for the shear viscosity requires the large $s$
limit of the integral over $s$ in Eq.\ (\ref{7.1}). It can be verified that $%
\Omega _{G}^{\eta }\left( s\right) $ has no invariant part, so that this
limit is expected to exist. This issue is discussed in some detail in
Appendix \ref{ap5}. Then the Green-Kubo expression for shear viscosity can
be written as
\begin{equation}
\eta =\Omega _{H}^{\eta }\left( 0\right) +\frac{1}{d^{2}+d-2} \sum_{i=1}^{d}
\sum_{j=1}^{d} \int d{\bm v}_{1}h_{ij}({\bm v}_{1})\mathcal{D}_{ij}\left({%
\bm v}_{1}\right),  \label{7.22}
\end{equation}
where $D_{ij}\left({\bm v}\right) $ is a solution to the integral equation
\begin{equation}
\left( \frac{\zeta_{0}}{2} \frac{\partial}{\partial {\bm v}} \cdot {\bm v} +
M + \frac{\zeta_{0}}{2}\right) \mathcal{D}_{ij}\left( {\bm v}\right) =\gamma
_{ij}\left( {\bm v}\right) .  \label{7.23}
\end{equation}
Upon writing the above equation, it has been taken into account that terms
involving spatial derivatives give a vanishing contribution to the
expression of the shear viscosity. This is the traditional form in which
expressions for transport coefficients are obtained from a Chapman Enskog
expansion of a normal solution to the kinetic equation governing the
dynamics of the system.

\subsection{Evaluation in the Granular Enskog Approximation}

The further neglect of velocity correlations in the collision operator $M$,
leads to the granular Enskog approximation, i.e., the results given by Eqs.\
(\ref{7.16a}) and (\ref{7.18}) apply with only the replacement $\Lambda$ by $%
\Lambda ^{E}$, with
\begin{equation}
\Lambda ^{E}(x) \equiv {\bm v} \cdot \frac{\partial}{\partial {\bm q}} +
M^{E} (x) + \frac{\zeta_{0}}{2} \frac{\partial}{\partial {\bm v}} \cdot {\bm %
v} ,  \label{7.25}
\end{equation}
and the operator $M^{E}$ given by,
\begin{eqnarray}
M^{E}X(x_{1}) &\equiv &-g_{h}^{2}\left( \sigma \right) \int dx_{2}\,
\overline{T} (x_{1},x_{2})\left[
f_{h}^{(1)}(v_{1})X(x_{2})+X(x_{1})f_{h}^{(1)}(v_{2})\right]  \notag \\
&&-\frac{1}{n_{h}} \int d {\bm q}_{2}\, c_{1}(x_{1,},{\bm q}_{2})\int d{\bm v%
}_{2}\, X(x_{2}).  \label{7.26}
\end{eqnarray}
The function $c_{1}$ is given in Eq.\ (\ref{5.2}). In the Enskog
approximation, the conjugate flux $\gamma_{ij}$ in Eq.\ (\ref{7.19a})
becomes
\begin{equation}
\gamma _{ij}\left( {\bm v}_{1}\right) =-\left( \Lambda^{E} +\frac{\zeta _{0}%
}{ 2}\right) \frac{1}{2} \left( q_{1,i} \frac{\partial}{\partial v_{1,j}}
+q_{1,j} \frac{\partial}{\partial v_{1,y}} -\frac{2}{d} \delta_{ij} {\bm q}%
_{1} \cdot \frac{\partial}{\partial {\bm v}_{1}} \right) f_{h}^{(1)}(v_{1}).
\label{7.27}
\end{equation}

One final simplification occurs for the shear viscosity and some other
transport coefficients. The mean field term in Eq.\ (\ref{7.26}) vanishes
when acting on $\gamma _{ij}\left( \mathbf{v}\right) $ and, therefore,
\begin{equation}
\left( \Lambda ^{E}+\frac{\zeta _{0}}{2}\right) \gamma _{ij}\left( {\bm v}%
\right) =\left( \mathcal{J}(x)+\frac{\zeta _{0}}{2}\right) \gamma
_{ij}\left( {\bm v}\right) ,  \label{7.31}
\end{equation}%
where the operator $\mathcal{J}(x)$ has been introduced,
\begin{equation}
\mathcal{J}(x)\equiv \frac{\zeta _{0}}{2}\frac{\partial }{\partial {\bm v}}%
\cdot {\bm v}-g_{h}^{(2)}(\sigma )\mathcal{I}.  \label{7.34a}
\end{equation}%
Here, $\mathcal{I}$ is the linearized Boltzmann collision operator for
inelastic hard spheres or disks,
\begin{equation}
\mathcal{I}X(x_{1})\equiv \int dx_{2}\overline{T}(x_{1},x_{2})\left[
f_{h}^{(1)}(v_{1})X(x_{2})+X(x_{1})f_{h}^{(1)}(v_{2})\right] .  \label{7.31a}
\end{equation}%
Therefore, the correlation function in Eq.\ (\ref{7.18})\ and the expression
for the shear viscosity in Eq.\ (\ref{7.22}) take the final forms
\begin{eqnarray}
\Omega _{G}^{\eta E}\left( s\right) &=&\frac{1}{d^{2}+d-2}\left[ 1+\frac{%
(1+\alpha )\sigma ^{d}n_{h}\pi ^{d/2}g_{h}^{(2)}(\sigma )}{4\Gamma \left(
\frac{d+4}{2}\right) }\right]  \notag \\
&&\times \sum_{i=1}^{d}\sum_{j=1}^{d}\int d{\bm v}\,v_{i}v_{j}\exp \left[
-s\left( \mathcal{J}+\frac{\zeta _{0}}{2}\right) \right] \gamma _{ij}({\bm v}%
)  \label{7.32}
\end{eqnarray}%
and
\begin{equation}
\eta =\Omega _{H}^{\eta E}(0)+\frac{1}{d^{2}+d-2}\sum_{i=1}^{d}%
\sum_{j=1}^{d}\int d{\bm v}\,h_{ij}({\bm v})\mathcal{D}_{ij}^{E}\left( {\bm v%
}\right) ,  \label{7.33}
\end{equation}%
respectively. The Enskog approximation for $\Omega _{H}^{\eta }\left(
0\right) $ is
\begin{equation}
\Omega _{H}^{\eta E}\left( 0\right) =\frac{\pi ^{(d-1)/2}(1+\alpha )\sigma
^{d+1}g_{h}^{(2)}(\sigma )}{2(d^{2}+2d)\Gamma \left( \frac{d+1}{2}\right) }%
\int d{\bm v}_{1}\int d{\bm v}_{2}\,|{\bm v}_{1}-{\bm v}_{2}|f_{h}^{\left(
1\right) }\left( v_{1}\right) f_{h}^{\left( 1\right) }\left( v_{2}\right)
\label{7.34}
\end{equation}%
and $\mathcal{D}_{ij}^{E}$ is a solution to the integral equation
\begin{equation}
\left( \mathcal{J}+\frac{\zeta _{0}}{2}\right) \mathcal{D}_{ij}^{E}({\bm v}%
)=\gamma _{ij}({\bm v}),  \label{7.35}
\end{equation}%
The above result for the shear viscosity agrees in detail with that obtained
in ref. \cite{GD99}, through a Chapman-Enskog procedure applied to the
non-linear granular Enskog equation.

The results in Eqs.\ (\ref{7.18})\ and (\ref{7.32})\ for the Green-Kubo
integrand are new. At the formally exact level, the integrand is given by
the correlation between the flux and the conjugate flux. In detail, the
contributions from $H_{ij}^{(2)}$ and the $\overline{T}(i,j)$ terms of $%
\overline{\mathcal{\ L}}$, appear to yield singularities at $t=0$, signaling
a possible nonanalytic dependence on $t$. This occurs even in the elastic
limit, and is a peculiarity of hard particle dynamics. Consequently,
previous theoretical and simulation studies have avoided this by studying
the Helfand forms for transport coefficients. The approximate kinetic theory
described here gives an explicit analytic estimate for $\Omega _{G}^{\eta
}\left( s\right) $, whose integral yields a good estimate for the transport
coefficients. This suggests that $\Omega _{G}^{\eta }\left( s\right) $ may
have a dominant analytic part with a relatively small non-analytic
correction.

\section{Discussion}

\label{s8}

Kinetic theory has been used extensively as a formal tool for approximate
evaluation of response functions, in the study of hard spheres as a model
for normal fluids. The objective of this work is to take a first step in the
development and application of this tool in the analogous field of granular
fluids. The advantage of developing kinetic theory in the context of linear
response functions, lies in the fact that the resulting theories are
inherently linear, and provide a more tractable setting to explore questions
such as aging to hydrodynamics and short wavelength behavior of the exact
hydrodynamic response. The two primary contributions here are: 1) the
development of a practical kinetic theory for an important class of granular
time correlation functions and, 2) the demonstration of its utility for the
evaluation of Helfand and Green-Kubo expressions for Navier-Stokes order
transport coefficients.

The linear kinetic theory is summarized by Eq.\ (\ref{4.15}). Based on
corresponding studies for the elastic limit of this equation, it is expected
to have a wide domain of validity with respect to space and time scales, as
well as densities. The nature of the approximation, short time functional
relationship, does not explicitly entail questions of inelasticity so it is
expected to apply as well for a finite range of inelasticity. It encompasses
the granular Boltzmann equation, in the low density limit, and the familiar
Enskog equation in the elastic limit. The focus here has been on
hydrodynamic response, but the theory includes hydrodynamics beyond the
Navier-Stokes approximation, and even describes very short wavelength
non-hydrodynamic behavior, that can be more important at moderate and high
densities for granular fluids. Finally, this kinetic theory applies beyond
the set of hydrodynamic fields considered here. For any observable, $%
z(\Gamma ;{\bm r})$, that can be written as a sum of single particle
functions so that
\begin{equation}
\widetilde{z}\left( \Gamma ,{\bm k}\right) =\sum_{r=1}^{N}e^{i{\bm k}\cdot {%
\bm q}_{r}}z\left( {\bm v}_{r}\right) ,  \label{8.1}
\end{equation}%
and for the same initial perturbation as considered in Sec. \ref{s2}, the
response functions are given in the appropriate units by
\begin{equation}
\widetilde{C}_{\gamma }\left( {\bm k};s\right) =\frac{1}{n_{h}}\int d{\bm v}%
\,z\left( {\bm v}\right) e^{-s\widetilde{\Lambda }({\bm k})}\widetilde{\psi }%
_{\gamma }^{(1)}({\bm v},{\bm k}).  \label{8.2}
\end{equation}%
This opens the possibility to study a wide range of experimental probes and
also fundamental questions such as the relationship between fluctuations and
response.

The second contribution, evaluation of the formal representations for
Navier-Stokes transport coefficients, begins the process of exploring the
utility of such formal representations, as well as verifying their
consistency with earlier Chapman-Enskog based studies. These are long
wavelength properties of the kinetic theory, and therefore a more controlled
context for its tests. For example, the Markovian approximation provides a
practical context for the introduction of velocity correlations associated
with the reference homogeneous state \cite{BGMyR04,BRMyG05}. Their effect on
transport coefficients at strong dissipation is expected to be important but
has not been quantified to date. The kinetic theory scheme developed here
also provides the basis for formulating and assessing more complex theories,
such as those describing mode coupling dynamical correlations, which are
expected to dominate at very high densities. Finally, the verification of
the agreement between the kinetic theory evaluation of the Helfand and
Green-Kubo representations here and the earlier Chapman-Enskog method
provides further support for the implicit assumptions of these complementary
formal approaches.

An interesting new result, both for normal and granular fluids, is the
expression of the Green-Kubo time-correlation function $\Omega _{G}(s)$. In
the Enskog approximation, the corresponding function for the shear viscosity
is given by Eq.\ (\ref{7.32}). To interpret this result, the time dependence
may be estimated from a leading order cumulant expansion,
\begin{equation}
\Omega _{G}^{\eta E}\left( s\right) \simeq \Omega _{G}^{\eta E}\left(
0\right) e^{-s/\tau },  \label{7.5A}
\end{equation}%
\begin{equation}
\frac{1}{\tau }=\frac{\sum_{i,j}^{d}\int d{\bm v}\,h_{ij}({\bm v})\left(
\mathcal{J}+\frac{\zeta _{0}}{2}\right) \gamma _{ij}\left( {\bm v}\right) }{%
\sum_{i,j}^{d}\int d{\bm v}\,h_{ij}\gamma _{ij}\left( {\bm v}\right) }\,.
\label{7.5B}
\end{equation}%
The corresponding Helfand correlation functions, $\Omega _{H}(s)$, can be
inferred directly from this,
\begin{equation}
\Omega _{H}^{\eta E}\left( s\right) \simeq \Omega _{H}^{\eta E}\left(
0\right) +\Omega _{G}^{\eta E}\left( 0\right) \tau \left( 1-e^{-s/\tau
}\right) ,  \label{3}
\end{equation}%
and the shear viscosity can be identified as
\begin{equation}
\eta \simeq \Omega _{H}^{\eta E}\left( 0\right) +\Omega _{G}^{\eta E}\left(
0\right) \tau .  \label{4}
\end{equation}%
These results expose the qualitative nature of the time dependence in each
case. The resulting shear viscosity in these approximations agrees with that
obtained by a leading order solution to the integral equation (\ref{7.35})
as an expansion in Sonine polynomials \cite{GD99}.

In conclusion, it is hoped that this work provides a starting point to
explore systematic analytic approximations to the hydrodynamic response of a
granular fluid, with the same attention to detail given in the context of
normal fluids. These, together with numerical studies of exact results,
provide a means to understand transport mechanisms in this system.

\section{Acknowledgements}

The research of A.B. and J.D. was supported in part by the Department of
Energy Grant (DE-FG03-98DP00218). A.B. also acknowledges a McGinty
Dissertation Fellowship and a IFT Michael J Harris Fellowship from the
University of Florida. The research of J.J.B. was partially supported by the
Ministerio de Educaci\'{o}n y Ciencia (Spain) through Grant No.
BFM2005-01398.

\appendix

\section{Local Homogeneous State}

\label{ap1}

The \emph{local} HCS ensemble chosen as the initial perturbation of the HCS,
represents a system decomposed into spatial cells, each in a HCS with its
own local temperature, density, and flow velocity. It is constructed
formally as follows. First, the HCS ensemble is determined as the solution
to the homogeneous, stationary Liouville equation (\ref{2.17}) in
dimensionless form,
\begin{equation}
\rho _{h}^{\ast }(\Gamma^{*})=\rho _{h}^{\ast }\left( \left\{ \frac{q_{rs}}{%
\ell} ,\frac{{\bm v}_{r}-{\bm U}_{h}}{v_{0}(T_{h})}; r,s=1,\ldots,N \right\}
\right) ,  \label{a.1}
\end{equation}
where ${\bm U}_{h}$, $T_{h}$, and $n_{h}$ (not shown explicitly) are the
flow field, temperature, and particle number density characterizing the HCS.
Next, a conservative external force is added to Eq.\ (\ref{2.12}) keeping
the same $\overline{\mathcal{L}}^{\ast }$ operator,
\begin{equation}
\left\{ \overline{\mathcal{L}}^{\ast }- \sum_{r=1}^{N}\left[ \frac{\partial}{%
\partial {\bm q}_{r}^{*}}\phi _{ext}^{*}\left({\bm q} _{r}\right) \right]
\cdot \frac{\partial}{\partial {\bm v^{*}_{r}}} \right\} \rho _{h}^{* \prime
}=0.  \label{a.2}
\end{equation}
Here $\phi _{ext}^{*} \equiv \phi_{ext}/2T_{h} $, with $\phi_{ext}({\bm r})$
being the potential associated with the external force. The solution of Eq.\
(\ref{a.2}) is, therefore, a function of this potential,
\begin{equation}
\rho _{h}^{\ast \prime }=\rho _{h}^{\ast \prime }\left( \left\{ \frac{q_{rs}%
}{\ell}, \frac{{\bm v}_{r}-{\bm U}_{h}}{v_{0}(t)}, \frac{\phi _{ext}\left({%
\bm q} _{r}\right) }{T_{h}} ; r,s=1,\ldots,N \right\} \right) .  \label{a.3}
\end{equation}
This can be considered as the nonuniform fluid ensemble corresponding to the
uniform limit $\rho _{h}^{\ast }$, since in general the density will be
nonuniform through its functional dependence on $\phi _{ext}\left({\bm r}
\right) $,
\begin{equation}
n=n\left[ {\bm r}| \phi_{ext} \right] ,  \label{a.4a}
\end{equation}
\begin{equation}
\phi_{ext}= \phi _{ext}\left[{\bm r}|n \right] .  \label{a.4b}
\end{equation}
The second equality assumes the functional dependence of the density on the
external potential is invertible so that the potential can be expressed as a
functional of the density field. For normal fluids in the equilibrium Gibbs
state, density functional theory assures that this is the case. In
particular, for any chosen density field there is a unique external
potential creating that field from the uniform state. It will be assumed
that these properties hold as well here for the granular fluid, so that Eq.\
(\ref{a.3}) can be expressed in terms of the local density instead of the
potential,
\begin{equation}
\rho _{h}^{\ast \prime }=\rho _{h}^{\ast \prime \prime }\left( \left\{ \frac{%
{q}_{rs}}{\ell}, \frac{{\bm v}_{r}-{\bm U}_{h}}{v_{0}(T_{h})}, n({\bm q}%
_{r})\ell ^{d}; r,s=1,\ldots,N \right\} \right) .  \label{a.5}
\end{equation}
With $\rho _{h}^{\ast \prime \prime }$ known from the solution to Eq.\ (\ref%
{a.2} ), the local HCS is constructed by the replacements
\begin{eqnarray}
\frac{{\bm v}_{r}-{\bm U}_{h}}{v_{0}(T_{h})} & \rightarrow & \frac{{\bm v}%
_{r}-{\bm U}_{h}- \delta {\bm U}({\bm q}_{r})}{v_{0}[T_{h}+\delta T ({\bm q}%
_{r})]},  \notag \\
n({\bm q}_{r}) & \rightarrow &n_{h}+ \delta n({\bm q}_{r}),  \label{a.6}
\end{eqnarray}
to get
\begin{equation}
\rho _{\ell h }^{\ast }\left[ \Gamma^{*} |\delta y^{*} \right] \equiv \rho
_{h}^{\ast \prime \prime }\left( \left\{ \frac{q_{rs}}{\ell}, \frac{{\bm v}%
_{r}-{\bm U}_{h}-\delta {\bm U}({\bm q}_{r})}{v_{0}\left[T_{h}+\delta T ({%
\bm q}_{r})\right]}, \left[ n_{h}+\delta n\left({\bm q}_{r}\right) \right]
\ell ^{d}; r,s=1, \ldots, N \right\} \right) .  \label{a.7}
\end{equation}

Note that the local HCS is no longer a solution to any Liouville equation,
but rather is simply a reference ensemble representing an hypothetical HCS
with different hydrodynamic parameters in each spatial cell of the fluid.
Its construction in the way presented above, supports that interpretation in
the sense that $\rho _{\ell h }^{\ast }\left[ \Gamma ^{*}|\delta y^{*}=\text{%
constant}\right] =\rho _{h}^{\ast }\left( \Gamma^{*}; y_{h}^{*}+\delta y
^{*}\right) $ so that both $\left\{ \rho _{\ell h}^{\ast }\left[
\Gamma^{*}|\delta y^{*} \right] \right\}_{\delta y^{*}=0} = \rho _{h}^{\ast
}\left(\Gamma^{*}; y_{h}^{*}\right) $ and all functional derivatives of $%
\rho _{\ell h }^{\ast }\left[ \Gamma^{*} |\delta y^{*} \right] $ become
derivatives of $\rho _{h}^{\ast }\left( \Gamma^{*}; y_{h}^{*}\right) $ at $%
\delta y^{*}=0$. More explicitly, it is
\begin{equation}
\int d{\bm r}_{1} \ldots \int d{\bm r}_{p}\, \left[ \frac{\delta^{p}
\rho_{lh}[\Gamma|\delta y]}{\delta y_{\alpha}({\bm r}_{1}) \ldots \delta
y_{\beta}({\bm r}_{p})} \right]_ {y =\left\{n_{h},T_{h},{\bm 0}\right\}} = %
\left[ \frac{\partial^{p} \rho_{h}(\Gamma; n_{h},T_{h},{\bm U}_{h})}{%
\partial y_{\alpha,h} \ldots \partial y_{\beta,h}} \right]_{{\bm U}_{h}={\bm %
0}},  \label{a.8}
\end{equation}
where $\Gamma$ is a point in the phase space associated to the original
positions and velocities and $\rho(\Gamma)$ the corresponding density. It is
instructive to carry out this construction of the local ensemble for the
case of a normal fluid. Then Eq.\ (\ref{a.1}) gives the familiar equilibrium
Gibbs ensemble, and Eq.\ (\ref{a.2}) gives the same ensemble with the
Hamiltonian modified to include the external potential. Finally, the
construction in Eq. (\ref{a.7}) gives the familiar local equilibrium
ensemble used in linear response theory for spatial perturbations of the
equilibrium state. In the grand ensemble the dependence on the local density
is implicit through a local chemical potential $\mu =\mu \left[{\bm r}| n %
\right] $.

\section{Two Particle Functional}

\label{ap2}

The reduced distribution functions associated with the solution to the
Liouville equation (\ref{2.20}) are defined as
\begin{equation}
f^{(m)}(x_{1},\ldots,x_{m},s)\equiv \frac{N!}{(N-m)!}\int dx_{m+1}\ldots
\int dx_{N}\, e^{- s \overline{\mathcal{L}}}\rho _{\ell h}[\Gamma | \delta
y].  \label{b.1}
\end{equation}
As is done in the main text, the asterisk indicating the use of
dimensionless variables is left implicit. The $f^{(m)}$'s are clearly not
independent functions. For example, $f^{(2)}(x_{1},x_{2},s)$ is related to $%
f^{(1)}(x_{1},s)$ by
\begin{equation}
(N-1)f^{(1)}(x_{1},s)=\int dx_{2}\, f^{(2)}(x_{1},x_{2},s).  \label{b.2}
\end{equation}
This implies that $f^{(2)}(x_{1},x_{2},s)$ has the representation
\begin{equation}
f^{(2)}(x_{1},x_{2},s)=f^{(1)}(x_{1},s)f^{(1)}(x_{2},s)g^{(2)}(x_{1},x_{2},s),
\label{b.3}
\end{equation}
where the pair correlation function $g^{(2)}(x_{1},x_{2},s)$ has the
properties
\begin{equation}
g^{(2)}(x_{1},x_{2},s)=g^{(2)}(x_{2},x_{1},s),  \label{b.4}
\end{equation}
\begin{equation}
\int dx_{2}f^{(1)}(x_{2},s)g^{(2)}(x_{1},x_{2},s)=\int
dx_{1}f^{(1)}(x_{1},s)g^{(2)}(x_{1},x_{2},s)=N-1.  \label{b.5}
\end{equation}
This in turn shows that $g^{(2)}(x_{1},x_{2},s)$ is a functional of $%
f^{(1)}(x_{1},s)$. Quite generally then, $f^{(2)}(x_{1},x_{2},s)$ can be
considered a functional of $f^{(1)}(x_{1},s)$,
\begin{equation}
f^{(2)}(x_{1},x_{2},s)=f^{(2)}\left[ x_{1},x_{2},s | f^{(1)}\left( s\right) %
\right].  \label{b.6}
\end{equation}
However, this functional relationship is not unique. The utility of Eq.\ (%
\ref{b.6}) lies in discovering a choice that leaves the simplest functional
dependence. In the low density limit, where it is expected that $%
g^{(2)}(x_{1},x_{2},s)$ tends to unity, this is clearly the case, since the
functional becomes a constant independent of $x_{1},x_{2},$ and $s$. More
generally, finding an appropriate functional form for $%
g^{(2)}(x_{1},x_{2},s) $ requires the detailed analysis of the full
many-body problem.

The corresponding functional for $\psi _{\gamma }^{(2)}(x_{1},x_{2},s)$
defined by Eq.\ (\ref{3.3a}) can be computed as
\begin{eqnarray}
\psi _{\gamma }^{(2)}(x_{1},x_{2},s) &\equiv &\left[ \frac{\delta
f^{(2)}(x_{1},x_{2},s)}{\delta y_{\gamma }({\bm0},0)}\right] _{\delta y=0}
\notag \\
&=&\int dx\,\left[ \frac{\delta f^{(2)}\left[ x_{1},x_{2},s|f^{(1)}\left(
s\right) \right] }{\delta f^{(1)}(x,s)}\,\frac{\delta f^{(1)}(x,s)}{\delta
y_{\gamma }({\bm0},0)}\right] _{\delta y=0}  \notag \\
&=&\int dx\,\left[ \frac{\delta f^{(2)}[x_{1},x_{2},s|f^{(1)}\left( s\right)
]}{\delta f^{(1)}(x,s)}\right] _{\delta y=0}\psi _{\gamma }^{(1)}(x,s)
\notag \\
&=&\int dx\,K(x_{1},x_{2},s;x)\psi _{\gamma }^{(1)}(x,s),  \label{b.7}
\end{eqnarray}%
with $K(x_{1},x_{2},s;x)$ given by Eq.\ (\ref{3.10}). This entails the
additional requirement that the functional $f^{(2)}[x_{1},x_{2},s|\cdot ]$
is independent of the specific initial fields $\delta y({\bm r},0)$, i.e. $%
f^{(2)}\left[ x_{1},x_{2},s|f^{(1)}\left( s\right) \right] $ depends on
these fields only through $f^{(1)}(x,s)$. Consequently, $K(x_{1},x_{2},s;x)$
is also independent of such initial data. This is expected, since the
collision operator $M(s)$ constructed from $K(x_{1},x_{2},s;x)$ by means of
Eq.\ (\ref{3.12}) should be universal for a wide class of initial
conditions. Finally, the functional form for $\psi _{\gamma
}^{(2)}(x_{1},x_{2},s)$ is seen to be linear in $\psi _{\gamma }^{(1)}(s)$,
while in general $f^{(2)}[x_{1},x_{2},s|f^{(1)}\left( s\right) ]$ is a
nonlinear functional of $f^{(1)}\left( s\right) $.

\section{Interpretation of $\widetilde{c}_{\protect\lambda }({\bm
v}, {\bm k%
})$}

\label{ap3}

The contribution to the action of the collision operator $\widetilde{M}({\bm %
k})$ on $\widetilde{\psi}_{\gamma}^{(1)}({\bm
v}_{1},-{\bm k},s) $ from the term proportional to $\widetilde{c}_{\lambda }(%
{\bm v}_{1},{\bm k}) $ in Eq.\ (\ref{4.14}) is
\begin{equation}
- \sum_{\lambda} \widetilde{c}_{\lambda }({\bm v}_{1},{\bm k}) \frac{1}{n_{h}%
} \int d {\bm v}\, a_{\lambda }\left({\bm v} \right) \widetilde{\psi }%
_{\gamma }^{(1)}( {\bm v},-{\bm k},s)=- \sum_{\lambda} \widetilde{c}%
_{\lambda }({\bm v}_{1},{\bm k })\widetilde{C}_{\lambda \gamma }\left({\bm k}%
;s\right) ,  \label{c.1}
\end{equation}
where Eq.\ (\ref{3.1}) has been employed. The above contribution depends on
only low order moments of the dependent variable $\widetilde{\psi}%
_{\gamma}^{(1)}$ in the kinetic equation (\ref{4.15}), and in fact only
those moments are of interest for determining the response functions. In
this sense, $\widetilde{c}_{\lambda}$ is a mean field operator rather than a
true collision operator, since its action does not depend directly on
differences in $\widetilde{\psi }_{\gamma }^{(1)}$ before and after a
collision like the first term of (\ref{4.14}). Instead, Eq. (\ref{4.11})
shows that $\widetilde{c}_{\lambda}$ reflects an average of collisional
effects induced through changes in the correlations. To provide some
interpretation of this term, consider first the elastic limit.

\subsection{Elastic limit}

In this case, $g_{\ell h }^{(2)}[x_{1},x_{2}|\delta y] = g_{\ell e }^{(2)}[{%
\bm q}_{1},{\bm q}_{2}|\delta n]$, independent of the velocities,
temperature, and flow field, i.e. it is a function of the spacial
coordinates and a functional of the density. Equation (\ref{4.11}) becomes
\begin{equation}
c_{\lambda }({\bm v}_{1},{\bm q}_{12})=\delta _{\lambda 1}n_{h}^{2}\int dx\,
\overline{T}(x_{1},x)\varphi \left( v_{1}\right) \varphi \left( v\right) %
\left[ \frac{\delta g_{\ell e }^{(2)}[{\bm q}_{1},{\bm q} |\delta n]}{\delta
n({\bm q}_{2})} \right] _{\delta n=0},  \label{c.3}
\end{equation}
where $\varphi \left( v \right) $ is the Maxwellian. The pair correlation
function for a nonuniform fluid, $g_{\ell e }^{(2)}[{\bm
q}_{1},{\bm q}_{2} |\delta n ]$, appears in the stationary first BBGKY
hierarchy equation ( \ref{3.3c}) in the presence of an external potential $%
\phi_{ext}$ associated with the given density (see Appendix \ref{ap1}),
\begin{equation*}
\left[ {\bm v}_{1} \cdot \frac{\partial}{\partial {\bm q}_{1}} - \left(
\frac{\partial}{\partial {\bm q}_{1}} \phi_{ext} \left[ {\bm q}_{1} |n %
\right] \right) \cdot \frac{\partial}{\partial {\bm v}_{1}} \right] n({\bm q}%
_{1}) \varphi (v_{1})
\end{equation*}
\begin{equation}
= \int dx_{2}\, \overline{ T}(x_{1},x_{2})\varphi \left( v_{1}\right)
\varphi \left( v_{2}\right) n\left( {\bm q}_{1}\right) n\left( {\bm q}%
_{2}\right) g_{\ell e}^{(2)} \left[ {\bm q} _{1},{\bm q}_{2} | \delta n%
\right] .  \label{c.4}
\end{equation}
The functional derivative of this equation with respect to $\delta n\left( {%
\bm q}_{3}\right) $ evaluated at at $\delta n =0$ gives
\begin{eqnarray}
n_{h} \varphi \left( v_{1}\right) {\bm v}_{1} \cdot \frac{\partial}{\partial
{\bm q}_{1}} C(q_{13}) &=&n_{h}\int dx_{2}\, \overline{T}(x_{1},x_{2})%
\varphi \left( v_{1}\right) \varphi \left( v_{2}\right) g_{e}^{(2)}\left(
q_{12} \right) \left[ \delta \left( {\bm q}_{31} \right) +\delta \left( {\bm %
q}_{32} \right) \right]  \notag \\
&&+n_{h}^{2}\int dx_{2}\, \overline{T}(x_{1},x_{2})\varphi \left(
v_{1}\right) \varphi \left( v_{2}\right) \left[ \frac{\delta g_{\ell e}^{(2)}%
\left[ {\bm q}_{1},{\bm q} _{2}| \delta n\right] }{\delta n\left( {\bm q}%
_{3}\right) }\right] _{\delta n=0}.  \label{c.5}
\end{eqnarray}
where it has been used that $g_{e}\left( q_{12} \right) =g_{\ell e }\left[ {%
\bm q}_{1},{\bm q}_{2} | 0\right] $ and $c\left( q_{13} \right) $ is the
direct correlation function defined by \cite{Hansen}
\begin{equation}
n_{h} c\left( {\bm q}_{1}-{\bm q}_{3}\right) =\delta \left( {\bm q}_{1}- {%
\bm q}_{3}\right) +2 n_{h} \left[ \frac{\delta \phi_{ext} \left[{\bm q}%
_{1}|n \right]}{\delta n({\bm q}_{3})} \right]_{\phi_{ext}=0}.  \label{c.6}
\end{equation}
The first term on the right hand side of Eq.\ (\ref{c.5}) can be evaluated
using the elastic limit of the explicit form for $\overline{T}(x_{1},x_{2})$
given in Eq.\ (\ref{2.13}). The result is
\begin{equation}
\int dx_{2}\, \overline{T}(x_{1},x_{2})\varphi \left( v_{1}\right) \varphi
\left( v_{2}\right) g_{e}^{(2)}\left( q_{12} \right) \left[ \delta \left( {%
\bm q}_{31} \right) +\delta \left( {\bm q}_{32} \right) \right] =g_{e}^{(2)}
\left( \sigma \right) \varphi \left( v_{1}\right) \delta (q_{13}-\sigma )
\widehat{\bm q}_{13} \cdot {\bm v}_{1}  \label{c.6a}
\end{equation}
Finally, therefore, Eq.\ (\ref{c.5} gives
\begin{equation*}
n_{h}\int dx_{2}\, \overline{T}(x_{1},x_{2})\varphi \left( v_{1}\right)
\varphi \left( v_{2}\right) \left[ \frac{\delta g_{\ell e}^{(2)} \left[ {\bm %
q}_{1},{\bm q} _{2}| \delta n\right] }{\delta n\left( {\bm q}_{3}\right) }%
\right]_{\delta n=0}
\end{equation*}
\begin{equation}
=\varphi \left( v_{1}\right) {\bm v}_{1} \cdot \frac{\partial}{\partial {\bm %
q}_{1}} \left[ c(q_{13})-g_{e}^{(2)}(\sigma) \Theta (q_{13}-\sigma) \right],
\label{c.9}
\end{equation}
where the delta function in Eq.\ (\ref{c.6}) has been written in terms of
the derivative of $\Theta (q_{13}-\sigma)$. Using this into Eq. (\ref{c.3})
gives the desired result,
\begin{equation}
c_{\lambda}({\bm v}_{1}, {\bm q}_{12})= \delta_{\lambda 1} n_{h} \varphi
(v_{1}) {\bm v}_{1} \cdot \frac{\partial}{\partial {\bm q}_{1}} \left[
c(q_{12})-g_{e}^{(2)}(\sigma) \Theta (q_{12}-\sigma) \right].  \label{c.9a}
\end{equation}
In the elastic case, it is seen that the contribution shown in Eq.\ (\ref%
{c.1}) is the same as that for an external force whose potential is $c\left(
q_{13} \right) -g_{h}^{(2)}\left( \sigma \right) \Theta \left( q_{13}-\sigma
\right) $. The direct correlation function has a discontinuity at $%
q_{13}=\sigma $, with value $c\left( \sigma \right) =g_{e}^{(2)}\left(
\sigma \right) $, so the subtracted theta function contribution assures that
this potential is continuous.

\subsection{Inelastic collisions}

For inelastic collisions, the effects of $\widetilde{c}_{\lambda }({\bm v }%
_{1,}{\bm k})$ are more complex and more difficult to interpret. However, a
significant difference from the elastic case can be seen already for the
simplest case of ${\bm k} ={\bm 0}$. From Eq.\ (\ref{4.10}), it follows that
\begin{equation}
\widetilde{c}_{\lambda }({\bm v}_{1,}{\bm 0})=\int d{\bm q}_{12}\,
c_{\lambda }(x_{1,}{\bm q}_{12})=\int dx\, \overline{T}%
(x_{1},x)f_{h}^{(1)}(v_{1})f_{h}^{(1)}(v) \left[ \frac{\partial
g_{h}^{(2)}(x_{1},x;n_{h},T_{h},{\bm U}_{h})}{\partial y_{\lambda,h }} %
\right]_{{\bm U}_{h}={\bm 0}},  \label{c.10}
\end{equation}
where use has been made of the identity
\begin{equation}
\int d{\bm q}\, \left[ \frac{\delta g_{\ell h }^{(2)}[x_{1},x_{2}|\delta y]}{%
\delta y_{\lambda }({\bm q})} \right]_{\delta y=0}= \left[ \frac{\partial
g_{h}^{(2)}(x_{1},x;y)}{\partial y_{\lambda }}\right]_{\{n,T,{\bm U}\}
=\{n_{h},T_{h},{\bm 0}\} }.  \label{c.11}
\end{equation}
This ${\bm k}={\bm 0}$ limit vanishes for elastic collisions, as it can be
seen directly from Eq. (\ref{c.9a}), but is nonzero for inelastic
collisions. This can be verified in the Enskog approximation, where velocity
correlations are neglected and, therefore,
\begin{eqnarray}
\widetilde{c}_{\lambda }^{E}({\bm v}_{1,}{\bm 0}) &= &\delta _{\lambda 1}
\left( \frac{\partial g_{h}^{(2)}(\sigma ;n )}{\partial n} \right)_{n=n_{h}}
\int dx\, \overline{ T}(x_{1},x)f_{h}^{(1)}(v_{1})f_{h}^{(1)}(v)  \notag \\
&=&\delta _{\lambda 1}\left( \frac{\partial \ln g_{h}^{(2)}(\sigma;n )}{%
\partial n} \right)_{n=n_{h}} \frac{\zeta_{0}}{2} \frac{\partial}{\partial {%
\bm v}_{1}} \cdot \left[ {\bm v}_{1} f_{h}^{(1)}(v_{1}) \right].
\label{c.12}
\end{eqnarray}
The second equality follows from the first hierarchy equation (\ref{3.3c})
particularized for the HCS,
\begin{equation}
\frac{\zeta_{0}}{2} \frac{\partial}{\partial {\bm v}_{1}}\cdot \left[ {\bm v}%
_{1} f_{h}^{(1)}(v_{1}) \right]=\int dx\overline{T}
(x_{1},x)g_{h}^{(2)}(x_{1},x)f_{h}^{(1)}(v_{1})f_{h}^{(1)}(V),  \label{c.13}
\end{equation}
when velocity correlations are neglected. Thus, it is seen that $\widetilde{%
c }_{\lambda }({\bm v}_{1,},{\bm 0})$ includes changes in the correlations
of collisional effects associated with cooling.

More generally, the Enskog approximation for arbitrary ${\bm k}$ reads
\begin{eqnarray}
\widetilde{c}_{\lambda }^{E}({\bm v}_{1},{\bm k}) & = & \delta _{\lambda
1}\int d{\bm q}\, e^{i{\bm k}\cdot {\bm q}}c_{\lambda }(x_{1},{\bm q})
\notag \\
& = & \int dx_{2}\, \overline{T}
(x_{1},x_{2})f_{h}^{(1)}(v_{1})f_{h}^{(1)}(v_{2})\int d{\bm q}\, e^{i {\bm k}%
\cdot {\bm q}} \left[ \frac{\delta g_{\ell h }^{(2)}[{\bm q}_{1},{\bm q}
_{2}|\delta y]}{\delta n({\bm q})}\right] _{\delta n=0}.  \label{c.14}
\end{eqnarray}

\section{Hydrodynamic modes of $\widetilde{\Lambda }(\mathbf{0})$}

\label{ap4}

In this Appendix, some of the details leading to the solution of the
eigenvalue problem (\ref{6.1}) at ${\bm k=0}$ are given. Consider first the
functions $\widetilde{\psi }_{\gamma }^{(1)}({\bm v}_{1},{\bm k})$, defined
in Eq.\ (\ref{4.17a}), at ${\bm k}=0$. By translational invariance%
\begin{equation}
\left[ \frac{\delta f_{\ell h}^{(1)}[\mathbf{q}_{1},\mathbf{v}_{1}|\delta y]%
}{\delta y_{\gamma }({\bm0})}\right] _{\delta y=0}=\left[ \frac{\delta
f_{\ell h}^{(1)}[\mathbf{q}_{1}+\mathbf{r},\mathbf{v}_{1}|\delta y]}{\delta
y_{\gamma }({\bm r})}\right] _{\delta y=0},  \label{d.0}
\end{equation}%
and so
\begin{eqnarray}
\widetilde{\psi }_{\gamma }^{(1)}({\bm v}_{1},{\bm0}) &=&\int d{\bm q}_{1}%
\left[ \frac{\delta f_{\ell h}^{(1)}[x_{1}|\delta y]}{\delta y_{\gamma }({\bm%
0})}\right] _{\delta y=0}=\frac{1}{V}\int d{\bm q}_{1}\int d{\bm r}\,\left[
\frac{\delta f_{\ell h}^{(1)}[x_{1}|\delta y]}{\delta y_{\gamma }({\bm r})}%
\right] _{\delta y=0}  \notag \\
&=&\left[ \frac{\partial f_{h}^{(1)}\left( {\bm v};y\right) }{\partial
y_{\gamma }}\right] _{y=y_{h}}.  \label{d.1}
\end{eqnarray}%
The last equality is a consequence of the construction of the local HCS,
assuring that all functional derivatives in the homogeneous limit are
related with ordinary derivatives of the HCS (see Appendix \ref{ap1}). Use
of the expression of the operator $\widetilde{\Lambda }$, Eq.\ (\ref{4.16}),
yields
\begin{equation}
\widetilde{\Lambda }({\bm0})\widetilde{\psi }_{\gamma }^{(1)}({\bm v}_{1},{%
\bm0})=\widetilde{M}\left( {\bm0}\right) \left[ \frac{\partial
f_{h}^{(1)}\left( {\bm v}_{1};y\right) }{\partial y_{\gamma }}\right]
_{y=y_{h}}+\frac{\zeta _{0}}{2}\frac{\partial }{\partial {\bm v}_{1}}\cdot
\left\{ {\bm v}_{1}\left[ \frac{\partial f_{h}^{(1)}\left( {\bm v}%
_{1};y\right) }{\partial y_{\gamma }}\right] _{y=y_{h}}\right\} .
\label{d.2}
\end{equation}%
Next, Eq.\ (\ref{4.14}) gives
\begin{eqnarray}
\widetilde{M}({\bm0})\left[ \frac{\partial f_{h}^{(1)}\left( {\bm v}%
_{1};y\right) }{\partial y_{\gamma }}\right] _{y=y_{h}} &=&-\int dx_{2}%
\overline{T}(x_{1},x_{2})g_{h}^{(2)}(x_{1},x_{2})\left\{ \frac{\partial }{%
\partial y_{\gamma }}\left[ f_{h}^{(1)}({\bm v}_{1};y)f_{h}^{(1)}({\bm v}%
_{2};y)\right] \right\} _{y=y_{h}}  \notag  \label{d.3} \\
&&-\frac{1}{n_{h}}\sum_{\lambda }\int dx\,\overline{T}%
(x_{1},x)f_{h}^{(1)}(v_{1})f_{h}^{(1)}(v)\left[ \frac{\partial
g_{h}^{(2)}(x_{1},x;\delta y)}{\partial y_{\lambda }}\right] _{y=y_{h}}
\notag \\
&&\times \int d{\bm v}_{2}\,a_{\lambda }({\bm v}_{2})\left[ \frac{\partial
f_{h}^{(1)}\left( {\bm v}_{2};y\right) }{\partial y_{\gamma }}\right]
_{y=y_{h}}  \notag \\
&=&-\left[ \frac{\partial }{\partial y_{\gamma }}\int dx_{2}\,\overline{T}%
(x_{1},x_{2})g_{h}^{(2)}(x_{1},x_{2};y)f_{h}^{(1)}({\bm v}_{1};y)f_{h}^{(1)}(%
{\bm v}_{2};y)\right] _{y=y_{h}}\,,  \notag \\
&&
\end{eqnarray}%
where Eq.\ (\ref{c.10}) has been employed. The right hand side in the above
equation can be further simplified by means of Eq.\ (\ref{c.13}), after
changing ${\bm v}_{r}$ into ${\bm v}_{r}-{\bm U}$,
\begin{equation}
\widetilde{M}\left( {\bm0}\right) \left[ \frac{\partial f_{h}^{(1)}\left( {%
\bm v}_{1};y\right) }{\partial y_{\gamma }}\right] _{y=y_{h}}=-\left\{ \frac{%
\partial }{\partial y_{\gamma }}\left[ \frac{\zeta _{0}(y)}{2}\frac{\partial
}{\partial {\bm v}_{1}}\cdot ({\bm v}_{1}-{\bm U})f_{h}^{(1)}({\bm v}_{1},y)%
\right] \right\} _{y=y_{h}},  \label{d.6}
\end{equation}%
and Eq.\ (\ref{d.2}) becomes
\begin{equation}
\widetilde{\Lambda }({\bm0})\widetilde{\psi }_{\gamma }^{(1)}({\bm v}_{1},{%
\bm0})=-\left[ \frac{\partial \zeta _{0}(y)}{\partial y_{\gamma }}\right]
_{y=y_{h}}\frac{1}{2}\frac{\partial }{\partial {\bm v}_{1}}\cdot \left[ {\bm %
v}_{1}f_{h}^{(1)}(v_{1})\right] +\frac{\zeta _{0}}{2}\left( \frac{\partial {%
\bm U}}{\partial y_{\gamma }}\right) _{y=y_{h}}\cdot \frac{\partial }{%
\partial {\bm v}_{1}}f_{h}^{(1)}(v_{1}).  \label{d.7}
\end{equation}%
Realizing that
\begin{equation}
\left[ \frac{\partial f_{h}^{(1)}\left( {\bm v},y\right) }{\partial y_{2}}%
\right] _{y=y_{h}}=-\frac{1}{2}\frac{\partial }{\partial {\bm v}}\cdot \left[
{\bm v}f_{h}^{(1)}\left( v\right) \right]  \label{d.8}
\end{equation}%
and
\begin{equation}
\left[ \frac{\partial f_{h}^{(1)}({\bm v},y)}{\partial {\bm y}_{3}}\right]
_{y=y_{h}}=-\frac{\partial f_{h}^{(1)}(v)}{\partial {\bm v}},  \label{d.8a}
\end{equation}%
Eq.\ (\ref{d.7}) is seen to be equivalent to
\begin{equation}
\widetilde{\Lambda }({\bm0})\widetilde{\psi }_{\gamma }^{(1)}({\bm v}_{1},{%
\bm0})=\left[ \frac{\partial \zeta _{0}(y)}{\partial y_{\gamma }}\right]
_{y=y_{h}}\widetilde{\psi }_{2}^{(1)}({\bm v}_{1},{\bm0})-\frac{\zeta _{0}}{2%
}\left( \frac{\partial {\bm U}}{\partial y_{\gamma }}\right) _{y=y_{h}}\cdot
\widetilde{\bm\psi }_{3}^{(1)}({\bm v}_{1},{\bm0}).  \label{d.8b}
\end{equation}%
Above, the index 3 and vectorial notation has been used to identify the $d$
components associated to the velocity field.

The specific cases following from Eq.\ (\ref{d.8b}) are
\begin{equation}
\widetilde{\Lambda }({\bm0})\widetilde{\psi }_{1}^{(1)}({\bm v}_{1},{\bm0}%
)=n_{h}\left( \frac{\partial \zeta _{0}}{\partial n_{h}}\right) _{T_{h}}%
\widetilde{\psi }_{2}^{(1)}({\bm v}_{1},{\bm0}),  \label{d.9}
\end{equation}%
\begin{equation}
\widetilde{\Lambda }({\bm0})\widetilde{\psi }_{2}^{(1)}({\bm v}_{1},{\bm0})=%
\frac{\zeta _{0}}{2}\widetilde{\psi }_{2}^{(1)}({\bm v}_{1},{\bm0}),
\label{d.10}
\end{equation}%
and
\begin{equation}
\widetilde{\Lambda }({\bm0})\widetilde{\bm\psi }_{3}^{(1)}({\bm v}_{1},{\bm0}%
)=-\frac{\zeta _{0}}{2}\widetilde{\bm\psi }_{3}^{(1)}({\bm v}_{1},{\bm0}).
\label{d.11}
\end{equation}%
Equations (\ref{d.9}) and (\ref{d.10}) can be combined to give
\begin{equation}
\widetilde{\Lambda }({\bm0})\left[ \widetilde{\psi }_{1}^{(1)}({\bm v}_{1},{%
\bm0})-2\left( \frac{\partial \ln \zeta _{0}}{\partial \ln n_{h}}\right)
_{T_{h}}\widetilde{\psi }_{2}^{(1)}({\bm v}_{1},{\bm0})\right] =0.
\label{d.12}
\end{equation}%
The above equations (\ref{d.10})-(\ref{d.12}) have the form of eigenvalue
equations
\begin{equation}
\widetilde{\Lambda }({\bm0})\phi ^{(\beta )}({\bm0},{\bm v}_{1})=\lambda
^{(\beta )}({\bm0})\phi ^{(\beta )}({\bm0},{\bm v}_{1}),  \label{d.13}
\end{equation}%
$\beta =1,\ldots ,d+2$, with the eigenvalues and eigenfunctions given in
Eqs.\ (\ref{6.2}) and (\ref{6.3}).

\section{Transport Coefficients}

\label{ap5}

In this appendix, the pressure, the cooling rate, and the Navier-Stokes
transport coefficients are evaluated in the kinetic theory approximations
developed in the main text. These quantities are identified from the
linearized phenomenological Navier-Stokes equations for an isolated granular
fluid, that in dimensionless units have the form
\begin{equation}
\frac{\partial}{\partial s} \delta \widetilde{y}^{*}_{\beta} ({\bm k}^{*},s)
+\sum_{\gamma=1}^{d+2} \mathcal{K}_{\beta \gamma}^{* hyd} \delta
y^{*}_{\gamma} ({\bm k}^{*},s) =0,  \label{E.0.1}
\end{equation}
where the transport matrix $\mathcal{K}^{* hyd}$ is found to be block
diagonal with a ``longitudinal'' part, corresponding to the fields $\{
\delta n^{*}, \delta T^{*}, \delta U_{\perp}^{*} \}$, given by
\begin{equation}
\mathcal{K}^{ \ast hyd }\left( {\bm k}^{\ast }\right) =\left(
\begin{array}{ccc}
0 & 0 & -ik^{\ast} \\
\zeta _{0}^{\ast }\frac{\partial \ln \zeta _{0}}{\partial \ln n_{h}}+\left(
\frac{2 \mu ^{\ast }}{d}-\zeta^{ \ast n}\right) k^{\ast 2} & \frac{\zeta
_{0}^{\ast }}{2}+\left( \frac{2 \lambda ^{\ast }}{d}-\zeta^{\ast T }\right)
k^{\ast 2} & - i \left( \frac{2 p_{h}^{\ast }}{d} +\zeta^{ \ast U }\right)
k^{\ast } \\
-\frac{i p_{h}^{\ast }}{2} \frac{\partial \ln p_{h}}{\partial \ln n_{h}}
k^{\ast } & -\frac{i p^{\ast}_{h}}{2}\, k^{\ast} & -\frac{\zeta _{0}^{\ast }%
}{2}+\left[ \frac{2(d-1)}{d}\eta ^{\ast }+\kappa ^{\ast }\right] k^{\ast 2}%
\end{array}
\right).  \label{E.0.2}
\end{equation}
The ``transverse'' components, $\delta U^{*}_{\perp,i}$, decouple from the
longitudinal ones, and their transport matrix reads
\begin{equation}
\mathcal{K}^{hyd *} ({\bm k}) =\left( - \frac{\zeta_{0}^{*}}{2} + \eta^{*}
k^{* 2} \right) I,  \label{E.0.3}
\end{equation}
with $I$ being the unit matrix of dimesnsion $d-1$. These expressions
include the unspecified functions $p_{h}(n_{h},T_{h})$ and $\zeta
_{h}\left(n_{h},T_{h} \right) $, defining the pressure and the cooling rate
of the HCS, respectively, as well as the unknown shear viscosity $\eta
(n_{h},T_{h})$, the bulk viscosity $\kappa (n_{h},T_{h})$, the thermal
conductivity $\lambda (n_{h},T_{h})$, and the new coefficient, $\mu
(n_{h},T_{h}) $, associated with the contribution of the density gradient to
the heat transport in a granular fluid. Finally, $\zeta^{n}(n_{h},T_{h})$, $%
\zeta^{T}(n_{h},T_{h})$, and $\zeta^{U}(n_{h},T_{h})$ are transport
coefficients arising from the local cooling rate. The dimensionless form of
the cooling rate was defined in Eq.\ (\ref{2.19}), and the definitions for
the remaining dimensionless quantities are
\begin{equation*}
p^{\ast }_{h}=\frac{p_{h}}{n_{h}T_{h}}, \quad \eta ^{\ast }=\frac{\eta}{
mn_{h}l v_{0}}, \quad \kappa ^{\ast }=\frac{\kappa}{mn_{h} l v_{0}},
\end{equation*}
\begin{equation*}
\lambda ^{\ast }=\frac{\lambda}{l n_{h} v_{0} }, \quad \mu ^{\ast }=\frac{\mu%
}{l T_{h} v_{0}},
\end{equation*}
\begin{equation}
\zeta^{\ast U }=\zeta^{U}, \quad \zeta^{\ast n }=\frac{n_{h} \zeta^{n}}{ l
v_{0}}, \quad \zeta^{ \ast T }=\frac{T_{h} \zeta^{T}}{ l v_{0}}.
\label{E.0.4}
\end{equation}
Formal expressions for these parameters have been obtained in ref. \cite%
{BDB06}, by carrying out a linear response analysis. Each of those
expressions is evaluated below by means of the kinetic theory developed in
the text, for comparison to the results obtained by a Chapman-Enskog
solution to the nonlinear Enskog equation \cite{GD99}. First, some general
considerations that apply to all of the transport coefficients will be
addressed. Dimensionless units will be used in the remaining of this
Appendix, and the asterisk will be suppressed for simplicity, as done in the
main text.

\subsection{Reduced Form of the Transport Coefficients}

The Helfand form for a generic dimensionless transport coefficient $\chi $
as obtained in ref.\ \cite{BDB06} is
\begin{equation}
\chi =\lim \Omega _{H}\left( s\right) =\lim \frac{1}{V}\int d\Gamma
\,j\left( \Gamma \right) \left( 1-\mathcal{P}\right) e^{-s\left( \overline{%
\mathcal{L}}-\lambda \right) }\mathcal{M}\left( \Gamma \right) ,
\label{E.1.1}
\end{equation}%
where $j\left( \Gamma \right) $ is a flux associated with the
densities of the hydrodynamic variables, and $\lambda $ denotes one
of the eigenvalues in (\ref{6.2}). These fluxes have the generic
form
\begin{equation}
j\left( \Gamma \right) =\sum_{r=1}^{N}j_{1}\left( x_{l}\right)
+\sum_{r=1}^{N}\sum_{s\neq r}^{N}j_{2}\left( x_{r},x_{s}\right) ,
\label{E.1.2}
\end{equation}%
$j_{1}$ and $j_{2}$ being one and two-particle functions of the phase point,
respectively. The adjoint functions $\mathcal{M}\left( \Gamma \right) $ are
related to the conjugate densities in the hydrodynamic response defined in
Eq.\ (\ref{2.27}), and read
\begin{equation}
\mathcal{M}\left( \Gamma \right) =M\int d{\bm r}\,\widehat{\bm k}\cdot {\bm r%
}\left[ \frac{\delta \rho _{lh}[\Gamma |y]}{\delta y\left( {\bm r}\right) }%
\right] _{y=y_{h}},  \label{E.1.3}
\end{equation}%
$M$ being some constant. Moreover, $\mathcal{P}$ is the projection operator
defined by
\begin{equation}
\mathcal{P}(\Gamma )X\left( \Gamma \right) =\frac{1}{V}\sum_{\gamma =1}^{d+2}%
\widetilde{\psi }_{\gamma }\left( \Gamma ;{\bm0}\right) \int d\Gamma
^{\prime }\,\widetilde{a}_{\gamma }\left( \Gamma ^{\prime };{\bm0}\right)
X\left( \Gamma ^{\prime }\right) ,  \label{E.1.4}
\end{equation}%
where the phase functions $\widetilde{a}_{\gamma }\left( \Gamma ;{\bm0}%
\right) $ and $\widetilde{\psi }_{\gamma }\left( \Gamma ;{\bm0}\right) $ are
the densities defined by Eqs.\ (\ref{2.22}) and the Fourier transform of (%
\ref{2.27}), respectively, both evaluated at ${\bm k}={\bm0}$. The generator
for the dynamics is $\overline{\mathcal{L}}-\lambda $, where $\lambda $ is
one of the hydrodynamic modes at ${\bm k}={\bm0}$ identified in Eq.\ (\ref%
{6.2}). Both the projection operator and the additional time dependence of
the term containing $\lambda $ are necessary to ensure that the long time
limit of the correlation function in Eq.\ (\ref{E.1.1}) is well defined.

Proceeding as in Sec. \ref{s7} for the shear viscosity, a reduced expression
for these transport coefficients in the Markov approximation can be
obtained. The generic correlation function $\Omega _{H}\left( s\right) $ of
Eq.\ (\ref{E.1.1}) becomes
\begin{equation}
\Omega _{H}\left( s\right) \simeq \frac{1}{V}\int dx_{1}\, J\left(
x_{1}\right) \left[ 1-\mathcal{P}^{\left( 1\right)} (x_{1})\right] e^{ -s
(\Lambda - \lambda)} \mathcal{M} ^{\left( 1\right) }\left( x_{1}\right).
\label{E.1.6}
\end{equation}
The direct flux $J\left( x_{1}\right) $ in this reduced time correlation
function is
\begin{equation}
J\left( x\right) \equiv j_{1}\left( x\right) +\int dx_{1}\int dx_{2}\,
j_{2}\left( x_{1},x_{2}\right) K\left( x_{1},x_{2},0;x\right),  \label{E.1.7}
\end{equation}
where $K\left( x_{1},x_{2},0;x\right) $ is the kernel for the collision
operator given by Eq.\ (\ref{4.8}), and $\mathcal{M} ^{\left( 1\right) }$ is
the one-particle function in the hierarchy associated with $\mathcal{M} $
\begin{equation}
\mathcal{M} ^{\left( 1\right) }\left( x_{1}\right) = N \int dx_{2} \ldots
\int dx_{N}\, \mathcal{M} \left( \Gamma \right) .  \label{E.1.8}
\end{equation}
The generator for the Markov dynamics $\Lambda $ is given in (\ref{4.5a}).
Lastly, $\mathcal{P}^{\left( 1\right) }$ is the one particle analog of $%
\mathcal{P}$ defined in Eq.\ (\ref{E.1.4}),
\begin{equation}
\mathcal{P}^{\left( 1\right) } (x_{1}) X\left( x_{1}\right) \equiv \frac{1}{V%
} \sum_{\gamma=1}^{d+2} \widetilde{\psi}_{\gamma }^{(1)}\left( {\bm v}_{1},{%
\bm 0}\right) \frac{1}{n_{h}}\int dx\, a_{\gamma }\left( {\bm v}\right)
X\left( x\right) ,  \label{E.1.10}
\end{equation}
where the $\widetilde{\psi}_{\gamma}^{(1)}({\bm v}_{1},{\bm 0})$ are the
functions defined in Eq.\ (\ref{d.1}).

Equation (\ref{E.1.1}) can be transformed into the Green-Kubo form to give
\begin{equation}
\chi =\Omega _{H}\left( 0\right) +\lim \int_{0}^{s}ds^{\prime }\, \frac{%
\partial}{\partial s^{\prime}} \Omega _{H}\left( s^{\prime }\right) \equiv
\Omega _{H}(0)+\lim \int_{0}^{s}ds^{\prime }\, \Omega _{G}\left( s^{\prime
}\right) .  \label{E.1.10a}
\end{equation}
The Markov approximation in this representation follows directly from Eq.\ (%
\ref{E.1.6}),
\begin{equation}
\Omega _{H}(0) \simeq \frac{1}{V}\int dx_{1}\, J\left( x_{1}\right) \left[
1- \mathcal{P}^{\left( 1\right)}(x_{1})\right] \mathcal{M} ^{\left( 1\right)
}\left( x_{1}\right) ,  \label{E.1.11}
\end{equation}
\begin{equation}
\Omega _{G}\left( s\right) \simeq \frac{1}{V}\int dx_{1}\, J\left(
x_{1}\right) \left[ 1-\mathcal{P}^{\left( 1\right) } (x_{1})\right] e^{-s
(\Lambda - \lambda)} \gamma \left( x_{1}\right) ,  \label{E.1.11a}
\end{equation}
with the conjugate flux $\gamma (x_{1}) $ given by
\begin{equation}
\gamma \left( x_{1}\right) \equiv - \left( \Lambda - \lambda \right)
\mathcal{M}^{\left( 1\right) }\left( x_{1}\right) .  \label{E.1.11b}
\end{equation}
In this Green-Kubo form, the role of the projection operator can be readily
interpreted as follows. Taking into account that $\mathcal{P}^{(1)}$
projects over the subspace spanned by the hydrodynamic eigenfunctions of $%
\widetilde{\Lambda}({\bm 0})$, the property
\begin{equation}
\left[ 1-\mathcal{P}^{\left( 1\right)}(x_{1}) \right] e^{ - s \left( \Lambda
-\lambda \right)} =\left[ 1-\mathcal{P} ^{\left( 1\right)} (x_{1}) \right]
e^{- s (\Lambda -\lambda )} \left[ 1-\mathcal{P}^{\left( 1\right)} (x_{1}) %
\right] ,
\end{equation}
is obtained. This shows that the presence of the projection operator in Eq.\
(\ref{E.1.11a}) ensures that the generator of dynamics $\exp [-s ( \Lambda
-\lambda )] $ acts on a function that is orthogonal to its invariants and,
therefore, has a well defined long time limit.

The above expressions can be specialized to the granular Enskog
approximation by replacing the Markovian kernel $K$ in Eq.\ (\ref{E.1.7})
with its Enskog approximation, and by replacing the linear operator $\Lambda
$ by $\Lambda ^{E}$, defined in Eq.\ (\ref{7.25}). Further, the reduced
conjugate functions in (\ref{E.1.8}) and Eqs.\ (\ref{E.1.10}) in the Enskog
approximation become
\begin{equation}
\left\{ \mathcal{M} _{\gamma}^{(1)} (x) \right\} \simeq \left\{ M \widehat{%
\bm k} \cdot {\bm q} f_{h}^{(1)}(v),\frac{M}{2} \widehat{\bm k} \cdot {\bm q}
\frac{\partial}{\partial {\bm v}} \cdot \left[ {\bm v} f_{h}^{(1)}(v)\right]
,-M \widehat{\bm k} \cdot {\bm q} \frac{\partial}{\partial {\bm v}}
f_{h}^{(1)}(v) \right\}
\end{equation}
and
\begin{equation}
\left\{ \psi_{\gamma}^{(1)} ({\bm v},{\bm 0}) \right\} \simeq \left\{
f_{h}^{(1)}(v),\frac{1}{2}\, \frac{\partial}{\partial {\bm v}} \cdot \left[ {%
\bm v} f_{h}^{(1)}(v)\right] ,- \frac{\partial}{\partial {\bm v}}
f_{h}^{(1)}(v) \right\},
\end{equation}
respectively. Now, $f_{h}^{(1)}(v)$ is the one-particle distribution of the
HCS obtained from the nonlinear Enskog kinetic equation. In the rest of the
appendix, explicit expressions for the hydrodynamic parameters are given in
the granular Enskog approximation.

\subsection{Evaluation in the Enskog Approximation}

The Helfand and Green-Kubo expressions for the Navier-Stokes transport
coefficients of a hard sphere granular fluid are reported in ref. \cite%
{BDB06}. From those expressions, the $N$ particle functions $j\left( \Gamma
\right) $ and $\mathcal{M} \left( \Gamma \right) $, together with the
eigenvalue $\lambda$ appearing in Eq.\ (\ref{E.1.1}) for each of the
transport coefficients, can be read off. Then, following the procedure
illustrated above and, in the context of the shear viscosity in the main
text, the results reported below are obtained. Attention is restricted to
the transport coefficients that were calculated in ref. \cite{GD99}, which
excludes the transport coefficients $\zeta ^{T}$ and $\zeta ^{n}$ from the
local cooling rate. Moreover, the parameter $\ell$ defining the length scale
is chosen such that $n_{h} \ell^{3} =1$ in the following, for the sake of
simplicity.

\subsubsection{ The pressure}

The expression for the pressure has been identified in \cite{BDB06} as
\begin{equation}
p=1+\frac{\left( 1+\alpha \right) \sigma }{2Vd}\int dx_{1} \int dx_{2}\,
\delta (q_{12} - \sigma) \Theta \left( - \widehat{\bm q}_{12}\cdot {\bm g}
_{12}\right) ( \widehat{\bm q}_{12} \cdot {\bm g}_{12})^{2} f_{h}^{(2)} ({%
\bm q}_{12}, {\bm v}_{1}, {\bm v}_{2} ).  \label{E.2.1a}
\end{equation}
This is the second moment of the normal component of the relative velocity
averaged over the two-particle distribution at contact. In the Enskog
approximation, all velocity correlations in the two-particle distribution
are neglected, and the above expression simplifies to
\begin{equation}
p=1+ \frac{\pi^{d/2} (1+ \alpha) \sigma^{d}}{ 2 \Gamma (d/2) d}\,
g_{h}^{(2)}(\sigma).  \label{E.2.2}
\end{equation}

\subsubsection{ The zeroth order cooling rate $\protect\zeta _{0}$}

The homogeneous dynamics in Eq.\ (\ref{E.0.1}) is determined entirely
determined by $\zeta _{0}$, for which the expression
\begin{equation}
\zeta _{0}=\frac{1-\alpha ^{2}}{2Vd} \int dx_{1}\int dx_{2}\, \delta (q_{12}
- \sigma) \Theta \left( \widehat{\bm q}_{12}\cdot {\bm g} _{12}\right) |
\widehat{\bm q}_{12} \cdot {\bm g}_{12}|^{3} f_{h}^{(2)} ({\bm q}_{12}, {\bm %
v}_{1}, {\bm v}_{2} )  \label{E.2.0a}
\end{equation}
was derived in \cite{BDB06}. When the Enskog approximation for the
two-particle distribution function is used, this simplifies to
\begin{eqnarray}
\zeta _{0} & = & \frac{ \pi^{(d-1)/2} \sigma^{d-1} g_{h}^{(2)}
(\sigma)(1-\alpha^{2})}{2 \Gamma \left( \frac{d+3}{2} \right) d} \int d{\bm v%
}_{1} \int d{\bm v}_{2}\, g_{12}^{3}f_{h}^{(1)}\left( v_{1}\right)
f_{h}^{(1)}\left(v_{2}\right)  \notag \\
& = & \frac{\Gamma \left( d/2 \right) (1- \alpha)}{\pi^{1/2} \Gamma \left(
\frac{d+3}{2} \right) \sigma }\, (p-1) \int d{\bm v}_{1} \int d{\bm v}_{2}\,
g_{12}^{3}f_{h}^{(1)}\left(v _{1}\right) f_{h}^{(1)}\left( v_{2}\right) .
\label{E.2.1}
\end{eqnarray}

\subsubsection{The Euler transport coefficient $\protect\zeta ^{U}$}

This is a new transport coefficient unique to granular fluids. Its
expression, as obtained in ref. \cite{BDB06}, reads
\begin{equation}
\zeta ^{U}=\lim \Omega _{H}^{\zeta ^{U}}\left( s\right) =\Omega _{H}^{\zeta
^{U}}(0)+\lim \int_{0}^{s}ds^{\prime }\,\Omega _{G}^{\zeta ^{U}}\left(
s^{\prime }\right) ,  \label{E.3.1}
\end{equation}%
where
\begin{equation}
\Omega _{H}^{\zeta ^{U}}\left( s\right) =V^{-1}\int d\Gamma \,W\left( \Gamma
\right) \left( 1-\mathcal{P}\right) e^{-s\left( \overline{\mathcal{L}}+\frac{%
\zeta _{0}}{2}\right) }\mathcal{M}_{\zeta ^{U}}(\Gamma )  \label{E.3.2}
\end{equation}%
and
\begin{equation}
\Omega _{G}^{\zeta ^{U}}\left( s\right) =V^{-1}\int d\Gamma \,W\left( \Gamma
\right) \left( 1-\mathcal{P}\right) e^{-s\left( \overline{\mathcal{L}}+\frac{%
\zeta _{0}}{2}\right) }\Upsilon _{\zeta ^{U}}(\Gamma ).  \label{E.3.4}
\end{equation}%
In the above expressions, $W\left( \Gamma \right) $ is the source term in
the microscopic energy balance equation that characterizes the dissipation
due to the inelastic collisions,
\begin{equation}
W\left( \Gamma \right) =\frac{1-\alpha ^{2}}{2d}\sum_{r=1}^{N}\sum_{s\neq
r}^{N}\delta (q_{rs}-\sigma )\Theta (-\widehat{\bm q}_{rs}\cdot {\bm g}%
_{rs})|\widehat{\bm q}_{rs}\cdot {\bm g}_{rs}|^{3},  \label{E.3.5}
\end{equation}%
and the conjugate density and flux are
\begin{equation}
\mathcal{M}_{\zeta ^{U}}(\Gamma )=-\sum_{r}^{N}{\bm q}_{r}\cdot \frac{%
\partial }{\partial {\bm v}_{r}}\rho _{h}(\Gamma ),  \label{E.3.5a}
\end{equation}%
\begin{equation}
\Upsilon _{\zeta ^{U}}(\Gamma )=-\left( \overline{\mathcal{L}}+\frac{\zeta
_{0}}{2}\right) \mathcal{M}_{\zeta ^{U}}(\Gamma ).  \label{E.3.6}
\end{equation}%
The Helfand form for this transport coefficient in the Enskog approximation
can be obtained by application of Eq.\ (\ref{E.1.6}) with the result
\begin{equation}
\Omega _{H}^{\zeta ^{U}}\left( s\right) \simeq V^{-1}\int dx\,J_{\zeta
^{U}}(x)\left[ 1-\mathcal{P}^{\left( 1\right) }(x)\right] e^{-s\left(
\Lambda ^{E}+\frac{\zeta _{0}}{2}\right) }\mathcal{M}_{\zeta ^{U}}^{\left(
1\right) }(x),  \label{E.3.8}
\end{equation}%
where
\begin{equation}
J_{\zeta ^{U}}\left( {\bm v}\right) =\frac{\pi ^{\frac{d-1}{2}%
}g_{h}^{(2)}(\sigma )\sigma ^{d-1}}{\Gamma \left( \frac{d+3}{2}\right) d}%
\int d{\bm v}_{1}\,|{\bm v}-{\bm v}_{1}|^{3}f_{h}^{(1)}\left( {v}_{1}\right)
,  \label{E.3.9}
\end{equation}%
\begin{equation}
\mathcal{M}_{\zeta _{U}}^{\left( 1\right) }(x_{1})=-{\bm q}_{1}\cdot \frac{%
\partial }{\partial {\bm v}_{1}}\widetilde{\psi }_{1}^{(1)}({\bm v}_{1},{\bm0%
}),  \label{E.3.10}
\end{equation}%
and $\Lambda ^{E}$ is the collision operator given in Eq.\ (\ref{7.25}). The
Green-Kubo form of this transport coefficient is determined by $\Omega
_{H}^{\zeta ^{U}}(0)$ and $\Omega _{G}^{\zeta ^{U}}\left( s\right) $. Direct
evaluation of the former gives \cite{BDB06}
\begin{equation}
\Omega _{H}^{\zeta ^{U}}(0)=-\frac{3}{d}\left( 1-\alpha \right) \left(
p-1\right) .  \label{E.3.11}
\end{equation}%
Here $p$ is the pressure in the Enksog approximation, Eq.\ (\ref{E.2.2}).
The time correlation function is (see Eq.\ (\ref{E.1.11a}))
\begin{equation}
\Omega _{G}^{\zeta {U}}\left( s\right) =V^{-1}\int dx\,J_{\zeta ^{U}}\left(
x\right) \left[ 1-\mathcal{P}^{\left( 1\right) }(x)\right] e^{-s\left(
\Lambda ^{E}+\frac{\zeta _{0}}{2}\right) }\gamma _{\zeta ^{U}}\left( x\right)
\label{E.3.12}
\end{equation}%
with the conjugate flux given by
\begin{equation}
\gamma _{\zeta ^{U}}\left( x\right) =-\left( \Lambda ^{E}+\frac{\zeta _{0}}{2%
}\right) \mathcal{M}_{\zeta ^{U}}^{\left( 1\right) }(x).  \label{E.3.12a}
\end{equation}%
Proceeding as in the case of the shear viscosity, it can be shown that
\begin{equation}
e^{-s\left( \Lambda ^{E}+\frac{\zeta _{0}}{2}\right) }\gamma _{\zeta
^{U}}\left( {\bm v}\right) =e^{-s\left( \mathcal{J}+\frac{\zeta _{0}}{2}%
\right) }\gamma _{\zeta ^{U}}\left( {\bm v}\right) ,  \label{E.3.12c}
\end{equation}%
where $\mathcal{J}$ is the operator defined in Eq.\thinspace\ (\ref{7.34a}).
The time correlation function in Eq.\ (\ref{E.3.12}) can be expressed as
\begin{equation}
\Omega _{G}^{\zeta ^{U}}\left( s\right) =\int d{\bm v}\,J_{\zeta ^{U}}\left(
{\bm v}\right) \left[ 1-\mathcal{P}^{\left( 1\right) }(x)\right] e^{-s\left(
\mathcal{J}+\frac{\zeta _{0}}{2}\right) }\gamma _{\zeta ^{U}}\left( {\bm v}%
\right) .  \label{E.3.12f}
\end{equation}%
Therefore, this Euler order transport coefficient is given in the Enskog
approximation by
\begin{equation}
\zeta ^{U}=-\frac{3}{d}\left( 1-\alpha \right) \left( p-1\right) +\int d{\bm %
v}\,J_{\zeta ^{U}}\left( {\bm v}\right) \left[ 1-\mathcal{P}^{\left(
1\right) }(x)\right] \mathcal{D}_{\zeta ^{U}}\left( {\bm v}\right) ,
\label{E.3.12g}
\end{equation}%
where $\mathcal{D}_{\zeta ^{U}}\left( {\bm v}\right) $ is the solution to
the integral equation
\begin{equation}
\left( \mathcal{J}+\frac{\zeta _{0}}{2}\right) \mathcal{D}_{\zeta
^{U}}\left( {\bm v}\right) =\left[ 1-\mathcal{P}^{\left( 1\right) }(x)\right]
\gamma _{\zeta ^{U}}\left( {\bm v}\right) .  \label{E.3.12h}
\end{equation}

\subsubsection{The bulk viscosity $\protect\kappa $}

The expression for the bulk viscosity reported in ref. \cite{BDB06} is
\begin{equation}
\kappa =\lim \Omega _{H}^{\kappa }\left( s\right) =\Omega _{H}^{\kappa
}(0)+\lim \int_{0}^{s}ds^{\prime }\Omega _{G}^{\kappa }\left( s^{\prime }\,
\right) ,  \label{E.4.1}
\end{equation}
where
\begin{equation}
\Omega _{H}^{\kappa }\left( s\right) =- \frac{1}{Vd^{2}}\int d\Gamma\, \text{%
tr } \mathsf{H} \left( 1- \mathcal{P} \right) e^{-s \left( \overline{%
\mathcal{L}}+\frac{\zeta _{0}}{2} \right) }\mathcal{M}_{\kappa } (\Gamma),
\label{E.4.2}
\end{equation}
and
\begin{equation}
\Omega _{G}^{\kappa }\left( s\right) =- \frac{1}{V d^{2}}\int d\Gamma\,
\text{tr } \mathsf{H} \left( 1- \mathcal{P} \right) e^{- s \left( \overline{%
\mathcal{L}}+\frac{\zeta _{0}}{2} \right) }\Upsilon _{\kappa } (\Gamma).
\label{E.4.5}
\end{equation}
The direct flux $\text{tr } \mathsf{H}$ is the trace of the momentum flux
given in Eq.\ (\ref{7.3a}), and the adjoint density is the same as that for $%
\zeta^{U}$, i.e.,
\begin{equation}
\mathcal{M}_{\kappa }(\Gamma)= \mathcal{M}_{\zeta^{U}} (\Gamma) =
-\sum_{r=1}^{N} {\bm q}_{r}\cdot \frac{\partial}{\partial {\bm v}_{r}} \rho
_{h}(\Gamma).  \label{E.4.6}
\end{equation}
Consequently, the conjugate flux is also the same, $\Upsilon _{\kappa
}=\Upsilon _{\zeta ^{U}},$ given by (\ref{E.3.6}). In the Enskog
approximation, the time correlation function $\Omega _{H}^{\kappa }\left(
s\right) $ becomes
\begin{equation}
\Omega _{H}^{\kappa }\left( s\right) \simeq V^{-1} \int dx\, J_{\kappa
}\left(x \right) \left[ 1-\mathcal{P}^{\left( 1\right) }(x)\right] e^{-s
\left( \Lambda ^{E}+\frac{\zeta _{0}}{2}\right) }\mathcal{M} _{\kappa
}^{\left( 1\right) }(x),  \label{E.4.7}
\end{equation}
with the direct flux given by
\begin{eqnarray}
J_{\kappa }\left( x_{1} \right) &=& - \frac{v_{1}^{2}}{d^{2}}+\frac{1 }{2d}%
\, \frac{\partial \ln g_{h}^{(2)}(\sigma)}{\partial n_{h}} \left( p-1\right)
\notag \\
&&- \frac{\left( 1+\alpha \right) \sigma g _{h}^{(2)} \left( \sigma \right)}{%
2d^{2}} \int dx_{2}\, \delta \left( q_{12}-\sigma \right) \Theta \left( -%
\widehat{\bm q}_{12}\cdot \mathbf{g} _{12}\right) \left| \widehat{{\bm q}}%
_{12}\cdot {\bm g}_{12}\right| ^{2}f_{h}^{(1)}\left( v_{2}\right)
\label{E.4.8}
\end{eqnarray}
and $\mathcal{M}_{\kappa }^{\left( 1\right) }(x)=\mathcal{M}_{\zeta
^{U}}^{\left( 1\right) }(x)$, given in Eq.\ (\ref{E.3.10}). The
instantaneous contribution $\Omega _{H}^{\kappa }(0)$ is simply related to $%
\Omega _{H}^{\eta }(0)$ in (\ref{7.34}) through \cite{BDB06}
\begin{equation}
\Omega _{H}^{\kappa }(0)=\frac{d+2}{d}\, \Omega _{H}^{\eta }(0).
\label{E.4.9}
\end{equation}
Furthermore, the correlation function $\Omega _{G}^{\kappa }(s)$ vanishes.
This can be seen as follows. The conjugate flux and the generator of
dynamics can be simplified as in the case of $\zeta ^{U}$ above, so that
they become independent of the coordinate ${\bm
q}$. Then, the direct flux above can be simplified to give
\begin{eqnarray}
J_{\kappa }\left( {\bm v}_{1}\right) & = & - \left[ 1+\frac{(1+\alpha)
\sigma^{d} \pi^{d/2} g_{h}^{(2)}(\sigma)}{ 2 \Gamma \left( \frac{d}{2}
\right)d} \right] \frac{v_{1}^{2}}{d^{2}}  \notag \\
&& - \left[ \frac{1}{2} \frac{\partial \ln g_{h}^{(2)} (\sigma)}{\partial
n_{h}} +1 \right] \frac{p-1}{d}.
\end{eqnarray}
Since $J_{\kappa }\left( {\bm v}\right) $ is orthogonal to the subspace
spanned by $1-\mathcal{P}^{\left( 1\right) } (x) $, $\Omega _{G}^{\kappa
}(s) $ vanishes. Therefore, the bulk viscosity in the Enskog approximation
is simply
\begin{eqnarray}
\kappa &=&\frac{d+2}{d}\, \Omega _{H}^{\eta }(0)= \frac{\pi^{(d-1)/2}
(1+\alpha) \sigma^{d+1} g_{h}^{(2)}(\sigma)}{2 d^{2} \Gamma \left( \frac{d+1%
}{2} \right)} \int d {\bm v}_{1} \int d {\bm v} _{2}\, | {\bm v}_{1}-{\bm v}%
_{2} | f_{h}^{\left( 1\right) }\left( v_{1}\right) f_{h}^{\left( 1\right)
}\left( v_{2}\right)  \notag \\
&=& \frac{\pi^{-1/2} \Gamma \left( d/2 \right) \sigma (p-1)}{\Gamma \left(
\frac{d+1}{2} \right) d} \int d {\bm v}_{1} \int d {\bm v} _{2}\, | {\bm v}%
_{1}-{\bm v}_{2} | f_{h}^{\left( 1\right) }\left( v_{1}\right) f_{h}^{\left(
1\right) }\left( v_{2}\right).  \label{E.4.11}
\end{eqnarray}

\subsubsection{The thermal conductivity $\protect\lambda $}

The thermal conductivity is expressed in ref. \cite{BDB06} as
\begin{equation}
\lambda =\lim \Omega _{H}^{\lambda }\left( s\right) =\Omega _{H}^{\lambda
}(0)+\lim \int_{0}^{s}ds^{\prime }\,\Omega _{G}^{\lambda }\left( s^{\prime
}\right) ,  \label{E.5.1}
\end{equation}%
where
\begin{equation}
\Omega _{H}^{\lambda }\left( s\right) =-(Vd)^{-1}\int d\Gamma \,{\bm S}%
(\Gamma )\cdot \left( 1-\mathcal{P}\right) e^{-s\left( \overline{\mathcal{L}}%
-\frac{\zeta _{0}}{2}\right) }\bm{\mathcal M}_{\lambda }(\Gamma )
\label{E.5.2}
\end{equation}%
and
\begin{equation}
\Omega _{G}^{\lambda }\left( s\right) =-(Vd)^{-1}\int d\Gamma \,{\bm S}%
(\Gamma )\cdot \left( 1-\mathcal{P}\right) e^{-s\left( \overline{\mathcal{L}}%
-\frac{\zeta _{0}}{2}\right) }{\bm\Upsilon }_{\lambda }(\Gamma ).
\label{E.5.4}
\end{equation}%
The direct flux ${\bm S}$ is the heat flux in the microscopic balance
equation for the energy density,
\begin{equation}
\mathbf{S}=\sum_{r=1}^{N}v_{r}^{2}{\bm v}_{r}+\sum_{r=1}^{N}\sum_{s\neq
r}^{N}{\bm s}_{2}\left( x_{r},x_{s}\right) ,  \label{E.5.5}
\end{equation}%
with
\begin{equation}
{\bm s}_{2}\left( x_{r},x_{s}\right) =\frac{(1+\alpha )\sigma }{2}\delta
(q_{rs}-\sigma )\Theta \left( -\widehat{\bm q}_{rs}\cdot {\bm g}_{rs}\right)
(\widehat{\bm q}_{rs}\cdot {\bm g}_{rs})^{2}\left( \widehat{\bm q}_{rs}\cdot
{\bm G}_{rs}\right) \widehat{\bm q}_{rs},  \label{E.5.6}
\end{equation}%
where ${\bm G}_{rs}\equiv ({\bm r}+{\bm v}_{s})/2$. The conjugate density in
Eq.\ (\ref{E.5.2}) is
\begin{equation}
\bm{\mathcal M}_{\lambda }(\Gamma )=-\frac{1}{2}\sum_{r=1}^{N}{\bm q}_{r}%
\frac{\partial }{\partial {\bm v}_{r}}\cdot \left[ {\bm v}_{r}\rho
_{h}(\Gamma )\right]   \label{E.5.7}
\end{equation}%
and the associated flux is
\begin{equation}
{\bm\Upsilon }_{\lambda }(\Gamma )=-\left( \overline{\mathcal{L}}-\frac{%
\zeta _{0}}{2}\right) \bm{\mathcal M}_{\lambda }(\Gamma ).  \label{E.5.8}
\end{equation}%
In the Enskog approximation, the time correlation function determining the
Helfand form of the thermal conductivity $\lambda $ is
\begin{equation}
\Omega _{H}^{\lambda }\left( s\right) \simeq V^{-1}\int dx\,{\bm J}_{\lambda
}\left( x\right) \cdot \left[ 1-\mathcal{P}^{\left( 1\right) }(x)\right]
e^{-s\left( \Lambda ^{E}-\frac{\zeta _{0}}{2}\right) }\bm{\mathcal M}%
_{\lambda }^{\left( 1\right) }(x),  \label{E.5.9}
\end{equation}%
where
\begin{eqnarray}
{\bm J}_{\lambda }\left( x_{1}\right)  &=&\frac{v_{1}^{2}{\bm v}_{1}}{d}-%
\frac{(1+\alpha )\sigma g_{h}^{(2)}(\sigma )}{d}\int dx_{2}\,\delta \left(
q_{12}-\sigma \right) \Theta \left( -\widehat{\bm q}_{12}\cdot {\bm g}%
_{12}\right)   \notag \\
&&\times (\widehat{\bm q}_{12}\cdot {\bm g}_{12})^{2}(\widehat{\bm q}%
_{12}\cdot {\bm G}_{12})\widehat{\bm q}_{12}f_{h}^{(1)}\left( v_{2}\right)
\label{E.5.10}
\end{eqnarray}%
and
\begin{equation}
\bm{\mathcal M}_{\lambda }^{\left( 1\right) }(x_{1})=-\frac{{\bm q}_{1}}{2}%
\frac{\partial }{\partial {\bm v}_{1}}\cdot \left[ {\bm v}%
_{1}f_{h}^{(1)}(v_{1})\right] .  \label{E.5.11}
\end{equation}%
The function ${J}_{\lambda }$ defined in Eq.\ (\ref{E.5.10}) differs from
the generic form given in (\ref{E.1.7}) by a velocity independent term which
does not contribute to $\Omega _{H}^{\lambda }$. The contribution from the
initial correlation function to the Green-Kubo form of this transport
coefficient is given by
\begin{eqnarray}
\Omega _{H}^{\lambda }(0) &=&\frac{\Gamma (\frac{d}{2})}{2\pi ^{1/2}\Gamma
\left( \frac{3+d}{2}\right) }\sigma \left( p-1\right) \int d{\bm v}_{1}\int d%
{\bm v}_{2}\,f_{h}^{(1)}(v_{1})f_{h}^{(2)}(v_{2})  \notag \\
&&\times \left[ g_{12}G_{12}^{2}+g_{12}(\widehat{\bm g}_{12}\cdot {\bm G}%
_{12})^{2}+\frac{1}{4}g_{12}^{3}+\frac{3}{2}\left( \widehat{\bm g}_{12}\cdot
{\bm G}_{12}\right) \right]
\end{eqnarray}%
while the time-dependent correlation function becomes
\begin{equation}
\Omega _{G}^{\lambda }\left( s\right) =-\left[ 1+\frac{3(p-1)}{d+2}\right]
\int d{\bm v}\,\frac{v^{2}{\bm v}}{d}\cdot \left[ 1-\mathcal{P}^{\left(
1\right) }({\bm v})\right] e^{-s\left( \mathcal{J}-\frac{\zeta _{0}}{2}%
\right) }{\bm\gamma }_{\lambda }({\bm v}).  \label{E.5.10b}
\end{equation}%
Proceeding as in the case of the shear viscosity in Sec. \ref{s7}, it is
found that
\begin{equation}
{\bm\gamma }_{\lambda }=\left( \Lambda ^{E}-\frac{\zeta _{0}}{2}\right)
\frac{1}{2}{\bm q}\frac{\partial }{\partial {\bm v}}\cdot \left[ {\bm v}%
f_{h}^{1}(v)\right] .  \label{E.5.11c}
\end{equation}%
Thus the thermal conductivity in the Enskog approximation is given by
\begin{equation}
\lambda =\Omega _{H}^{\lambda }(0)-\left[ 1+\frac{3(p-1)}{d+2}\right] \int d{%
\bm v}\,\frac{v^{2}{\bm v}}{d}\cdot \left[ 1-\mathcal{P}^{\left( 1\right) }({%
\bm v})\right] \bm{\mathcal A}\left( {\bm v}\right) ,
\end{equation}%
where $\mathcal{\bm A}$ is the solution to the integral equation
\begin{equation}
\left( \mathcal{J}-\frac{\zeta _{0}}{2}\right) \bm{\mathcal A}({\bm v})=%
\left[ 1-\mathcal{P}^{(1)}({\bm v})\right] {\bm\gamma }_{\lambda }({\bm v}),
\label{E.5.12}
\end{equation}%
with $\mathcal{J}$ the operator defined in Eq. (\ref{7.34a}).

\subsubsection{The coefficient $\protect\mu $}

This is a new transport mechanism for granular fluids that arises due to the
inelasticity of collisions. As discussed in detail in ref. \cite{BDB06},
this transport coefficient consists of two time correlation functions, one
of which can be recognized as the time correlation part of the thermal
conductivity. Therefore, the quantity in terms of which the time correlation
function takes the simplest form is the linear combination
\begin{equation}
\overline{\mu }\equiv \mu -2\frac{\partial \ln \zeta _{h}}{\partial \ln n_{h}%
}\lambda =\lim \Omega _{H}^{\overline{\mu }}\left( s\right) =\Omega _{H}^{%
\overline{\mu }}(0)+\lim \int_{0}^{s}ds^{\prime }\,\Omega _{G}^{\overline{%
\mu }}\left( s^{\prime }\right) ,  \label{E.6.1}
\end{equation}%
with
\begin{equation}
\Omega _{H}^{\overline{\mu }}\left( s\right) =-(Vd)^{-1}\int d\Gamma \,{\bm S%
}(\Gamma )\cdot \left( 1-\mathcal{P}\right) e^{-s\overline{\mathcal{L}}}%
\bm{\mathcal M}_{\overline{\mu }}(\Gamma ),  \label{E.6.2}
\end{equation}%
and
\begin{equation}
\Omega _{G}^{\overline{\mu }}\left( s\right) =-(Vd)^{-1}\int d\Gamma \,{\bm S%
}(\Gamma )\cdot \left( 1-\mathcal{P}\right) e^{-s\overline{\mathcal{L}}}{\bm%
\Upsilon }_{\mu }(\Gamma ).  \label{E.6.4}
\end{equation}%
The direct flux ${\bm S}$ above is the heat flux given in Eq.\ (\ref{E.5.5}%
). The conjugate density is
\begin{eqnarray}
\bm{\mathcal M}_{\overline{\mu }} &=&\int d{\bm r}\,{\bm r}\left[ \frac{%
\delta \rho _{lh}}{\delta n\left( {\bm r}\right) }-2\frac{\partial \ln \zeta
_{0}}{\partial \ln n}T\frac{\delta \rho _{lh}}{\delta T\left( {\bm r}\right)
}\right] _{y=y_{h}}  \notag \\
&=&\int d{\bm r}\,{\bm r}\left[ \left( \frac{\delta \rho _{lh}[\Gamma |y]}{%
\delta n({\bm r})}\right) _{\zeta _{0}}\right] _{y=y_{h}},  \label{E.6.5}
\end{eqnarray}%
where the functional derivative with respect to the density in the last
equality is to be taken at constant cooling rate, as indicated. In the
Enskog approximation, the Helfand form becomes
\begin{equation}
\Omega _{H}^{\overline{\mu }}\left( s\right) \simeq V^{-1}\int dx\,{\bm J}_{%
\overline{\mu }}\left( x\right) \cdot \left[ 1-\mathcal{P}^{\left( 1\right)
}(x)\right] e^{-s\left( \Lambda ^{E}-\frac{\zeta _{0}}{2}\right) }%
\bm{\mathcal M}_{\overline{\mu }}^{\left( 1\right) }(x).  \label{E.6.6}
\end{equation}%
Here, ${\bm J}_{\overline{\mu }}\left( x\right) ={\bm J}_{\lambda
}\left( x\right) $, given in Eq.\ (\ref{E.5.10}), and
\begin{equation}
\bm{\mathcal M}_{\overline{\mu }}^{\left( 1\right) }={\bm q}\left\{
f_{h}^{(1)}\left( v\right) +\frac{\partial \ln \zeta _{0}}{\partial \ln n_{h}%
}\frac{\partial }{\partial {\bm v}}\cdot \left[ {\bm v}f_{h}^{(1)}(v)\right]
\right\} .  \label{E.6.7}
\end{equation}%
The Green-Kubo form is determined from $\Omega _{H}^{\overline{\mu }}(0)$
and $\Omega _{G}^{\overline{\mu }}\left( s\right) $. Direct evaluation shows
the relationship
\begin{equation}
\Omega _{H}^{\overline{\mu }}(0)=-2\frac{\partial \ln \zeta _{0}}{\partial
\ln n_{h}}\,\Omega _{H}^{\lambda }(0).  \label{E.6.8}
\end{equation}%
The time correlation function $\Omega _{G}^{\overline{\mu }}\left( s\right) $
can be simplified as in the previous cases yielding
\begin{equation}
\Omega _{G}^{\mu }\left( s\right) =\int d{\bm v}\,{\bm J}_{\overline{\mu }%
}\left( {\bm v}\right) \cdot \left[ 1-\mathcal{P}^{(1)}({\bm v})\right] e^{-s%
\mathcal{J}}{\bm\gamma }_{\overline{\mu }}\left( {\bm v}\right) ,
\label{E.6.9a}
\end{equation}%
with
\begin{equation}
{\bm\gamma }_{\mu ,i}\left( {\bm v}\right) =-{v}_{i}f_{h}^{(1)}\left(
v\right) -\left[ 1+\frac{1}{2}\frac{\partial \ln g_{h}^{(2)}\left( \sigma
\right) }{\partial \ln n_{h}}\right] Q_{i}\left[ f_{h}^{(1)}\left( v\right) %
\right] -2\frac{\partial \ln \zeta _{0}}{\partial \ln n_{h}}\gamma _{\lambda
,i}\,.  \label{E.6.10}
\end{equation}%
The Green-Kubo form for the transport coefficient $\mu $ is
\begin{equation}
\mu -2\frac{\partial \ln \zeta _{0}}{\partial \ln n_{h}}\lambda =-2\frac{%
\partial \ln \zeta _{0}}{\partial \ln n_{h}}\,\Omega _{H}^{\lambda }(0)+\int
d{\bm v}{\bm J}_{\lambda }\left( {\bm v}\right) \cdot \left[ 1-\mathcal{P}%
^{\left( 1\right) }(x)\right] \bm{\mathcal C}\left( {\bm v}\right) ,
\label{E.6.11}
\end{equation}%
where ${\bm C}\left( {\bm v}\right) $ is a solution to the integral equation
\begin{equation}
\mathcal{J}\bm {\mathcal C}({\bm v})=\left[ 1-\mathcal{P}^{(1)}(x)\right] {%
\bm J}_{\lambda }({\bm v}).  \label{E.6.12}
\end{equation}

All of the above results agree in detail (for $d=3$) with those obtained in
ref. \cite{GD99} by means of the Chapman-Enskog solution of the non-linear
Enskog kinetic equation.


\begin{thebibliography}{99}
\bibitem{chemE} Hydrodynamics derived from inelastic hard sphere models has
found widespread use in multi-phase computational fluid dynamics methods, to
describe numerous industrial processes involving solid particulates in gas
fluidized systems (e.g., software packages such as Fluent 2122 and MFIX).
\end{thebibliography}
\end{document}